\documentclass[journal=ancac3,manuscript=article,layout=twocolumn,9pt]{achemso}
\usepackage{extsizes}

\setkeys{acs}{maxauthors = 0}

\usepackage{graphicx}
\usepackage{braket}

\usepackage{color}
\usepackage{colortbl}
\usepackage{subfigure}

\usepackage{amsmath} 
\usepackage{amssymb} 
\usepackage{ulem} 
\usepackage[dvipsnames]{xcolor} 
\usepackage[version=3]{mhchem} 

\usepackage{bm}

\newcommand{\onlinecite}[1]{\hspace{-1 ex} \nocite{#1}\citenum{#1}} 

\renewcommand{\latin}{\textit}

\def\be{\begin{equation}}
\def\ee{\end{equation}}
\def\bea{\begin{eqnarray}}
\def\eea{\end{eqnarray}}

\let\oldmaketitle\maketitle
\let\maketitle\relax

\title{ Electron and hole $g$-factors and spin dynamics of negatively charged excitons in CdSe/CdS colloidal nanoplatelets with thick shells}

\author{E. V. Shornikova}
\affiliation{\footnotesize Experimentelle Physik 2, Technische Universit{\"a}t Dortmund, 44221 Dortmund, Germany}
\alsoaffiliation{\footnotesize Rzhanov Institute of Semiconductor Physics, Siberian Branch of Russian Academy of Sciences, 630090 Novosibirsk, Russia}
\email{elena.kozhemyakina@tu-dortmund.de}
\author{L. Biadala}
\affiliation{\footnotesize Institut d'Electronique, de Micro{\'e}lectronique et de Nanotechnologie, CNRS, 59652 Villeneuve-d'Ascq, France}
\email{louis.biadala@isen.iemn.univ-lille1.fr}
\author{D. R. Yakovlev}
\affiliation{\footnotesize Experimentelle Physik 2, Technische Universit{\"a}t Dortmund, 44221 Dortmund, Germany}
\email{dmitri.yakovlev@tu-dortmund.de}
\alsoaffiliation{\footnotesize Ioffe  Institute, Russian Academy of Sciences, 194021 St. Petersburg, Russia}
\author{D. H. Feng}
\affiliation{\footnotesize Experimentelle Physik 2, Technische Universit{\"a}t Dortmund, 44221 Dortmund, Germany}
\alsoaffiliation{State Key Laboratory of Precision Spectroscopy, East China Normal University, Shanghai 200062, China}
\author{V. F. Sapega}
\affiliation{\footnotesize Ioffe  Institute, Russian Academy of Sciences, 194021 St. Petersburg, Russia}
\author{N. Flipo}
\affiliation{\footnotesize Experimentelle Physik 2, Technische Universit{\"a}t Dortmund, 44221 Dortmund, Germany}
\author{A. A. Golovatenko}
\affiliation{\footnotesize Ioffe  Institute, Russian Academy of Sciences, 194021 St. Petersburg, Russia}
\author{M. A. Semina}
\affiliation{\footnotesize Ioffe  Institute, Russian Academy of Sciences, 194021 St. Petersburg, Russia}
\author{A. V. Rodina}
\affiliation{\footnotesize Ioffe  Institute, Russian Academy of Sciences, 194021 St. Petersburg, Russia}
\author{A. A. Mitioglu}
\affiliation{\footnotesize High Field Magnet Laboratory (HFML-EMFL), Radboud University, 6525 ED Nijmegen, The Netherlands}
\author{M. V. Ballottin}
\affiliation{\footnotesize High Field Magnet Laboratory (HFML-EMFL), Radboud University, 6525 ED Nijmegen, The Netherlands}
\author{P. C. M. Christianen}
\affiliation{\footnotesize High Field Magnet Laboratory (HFML-EMFL), Radboud University, 6525 ED Nijmegen, The Netherlands}
\author{Yu. G. Kusrayev}
\affiliation{\footnotesize Ioffe  Institute, Russian Academy of Sciences, 194021 St. Petersburg, Russia}
\author{M. Nasilowski}
\affiliation{\footnotesize Laboratoire de Physique et d'Etude des Mat\'{e}riaux, ESPCI, CNRS, 75231 Paris, France}
\author{B. Dubertret}
\affiliation{\footnotesize Laboratoire de Physique et d'Etude des Mat\'{e}riaux, ESPCI, CNRS, 75231 Paris, France}
\author{M. Bayer}
\affiliation{\footnotesize Experimentelle Physik 2, Technische Universit{\"a}t Dortmund, 44221 Dortmund, Germany}
\alsoaffiliation{\footnotesize Ioffe  Institute, Russian Academy of Sciences, 194021 St. Petersburg, Russia}

\begin{document}
\twocolumn[
\begin{@twocolumnfalse}
	\oldmaketitle
	\begin{abstract}
		We address spin properties and spin dynamics of carriers and charged excitons in CdSe/CdS colloidal nanoplatelets with thick shells. Magneto-optical studies are performed by time-resolved and polarization-resolved photoluminescence, spin-flip Raman scattering and picosecond pump-probe Faraday rotation in magnetic fields up to 30~T. We show that at low temperatures the nanoplatelets are negatively charged so that their photoluminescence is dominated by radiative recombination of negatively charged excitons (trions). Electron $g$-factor of 1.68 is measured and heavy-hole $g$-factor varying with increasing magnetic field from $-0.4$ to $-0.7$ is evaluated. Hole $g$-factors for two-dimensional structures are calculated  for various hole confining potentials for cubic- and wurtzite lattice in CdSe core. These calculations are extended for various quantum dots and nanoplatelets based on II-VI semiconductors. We developed a magneto-optical technique for the quantitative  evaluation of the nanoplatelets orientation in ensemble.
	\end{abstract}
\bigskip
\end{@twocolumnfalse}
]

Colloidal two-dimensional nanoplatelets (NPLs) are novel semiconductor nanostructures with remarkable optical properties due to atomically-controlled thickness.\cite{Ithurria2008,Nasilowski2016}
Most important, they have narrow ensemble emission spectra. At room temperature, the linewidth of ensemble photoluminescence (PL) spectrum does not exceed the one of a single nanocrystal, both for bare core\cite{Tessier2012} and core/shell\cite{Tessier2013nl} NPLs.
The nanoplatelets demonstrate outstanding properties such as high sensitivity to chemical agents due to vast reactive surfaces,\cite{Lorenzon2015} an ultralow stimulated emission threshold,\cite{Grim2014,Diroll2017} a very efficient fluorescence resonance energy transfer,\cite{Rowland2015} as well as a highly efficient charge carrier multiplication\cite{Aerts2014}. A plethora of materials can be synthesized as NPLs (organic led halide perovskites,\cite{Sichert2015,Dou2015} PbS,\cite{Schliehe2010,Dogan2015} PbSe,\cite{Koh2017} Cu$_{2-x}$S\cite{Sigman2003, Zhang2005}, GeS and GeSe,\cite{VaughnII2010} CdTe\cite{Ithurria2011nm}, CdS,\cite{Ithurria2011nm, Li2012} ZnS\cite{Bouet2014}, and HgTe\cite{Izquierdo2016}, see for review Ref.~\onlinecite{Nasilowski2016}) which allows for tuning the PL emission from the blue to the infrared spectral range.

Core/shell, and especially thick-shell nanocrystals (NCs), are at the center of intense research since 2008, when the first synthesis of core thick-shell CdSe/CdS quantum dots (QDs) has been reported.\cite{Mahler2008,Chen2008} These structures have been proven to exhibit suppressed blinking\cite{Mahler2008,Chen2008} and Auger recombination,\cite{GarciaSantamaria2009,Spinicelli2009,Htoon2010} and  almost $100\%$ quantum
yield.\cite{Javaux2013,Nasilowski2015} Photocharging in core/shell nanostructures is efficient and much more controllable compared to the bare ones. It was shown that at low temperatures dominant part of CdSe/CdS QDs is singly charged under illumination if the shell thickness exceeds 5~nm.\cite{Javaux2013,Liu2013} As a result, photoluminescence is dominated by recombination of the negatively charged excitons (trions), which ground state is bright and recombines within few nanoseconds. This allows to bypass the long-standing problem for colloidal NCs with large bright-dark splitting of exciton states and the dark exciton being the ground state. It also allows optical doping of colloidal NCs, providing resident carriers, which spin states can be addressed, controlled and manipulated by different means.

In this paper we address spin physics in CdSe/CdS colloidal nanoplatelets with thick shells by
means of time-resolved and polarization-resolved photoluminescence, spin-flip Raman scattering and
picosecond pump-probe Faraday rotation. High magnetic fields up to 30~T are used. Dominant role of
the negatively charged excitons (trions) in photoluminescence is found at cryogenic temperatures.
Trion spin dynamics is measured and electron and heavy-hole $g$-factors are evaluated. Hole
$g$-factors for two-dimensional structures are calculated to highlight the role of confining
potential and cubic- or wurtzite lattice in CdSe core. Magneto-optical technique for the
quantitative  evaluation of the nanoplatelets orientation in ensemble measurements is developed.

\section{Samples and optical spectra.}
\begin{figure*}[h!]
	\includegraphics{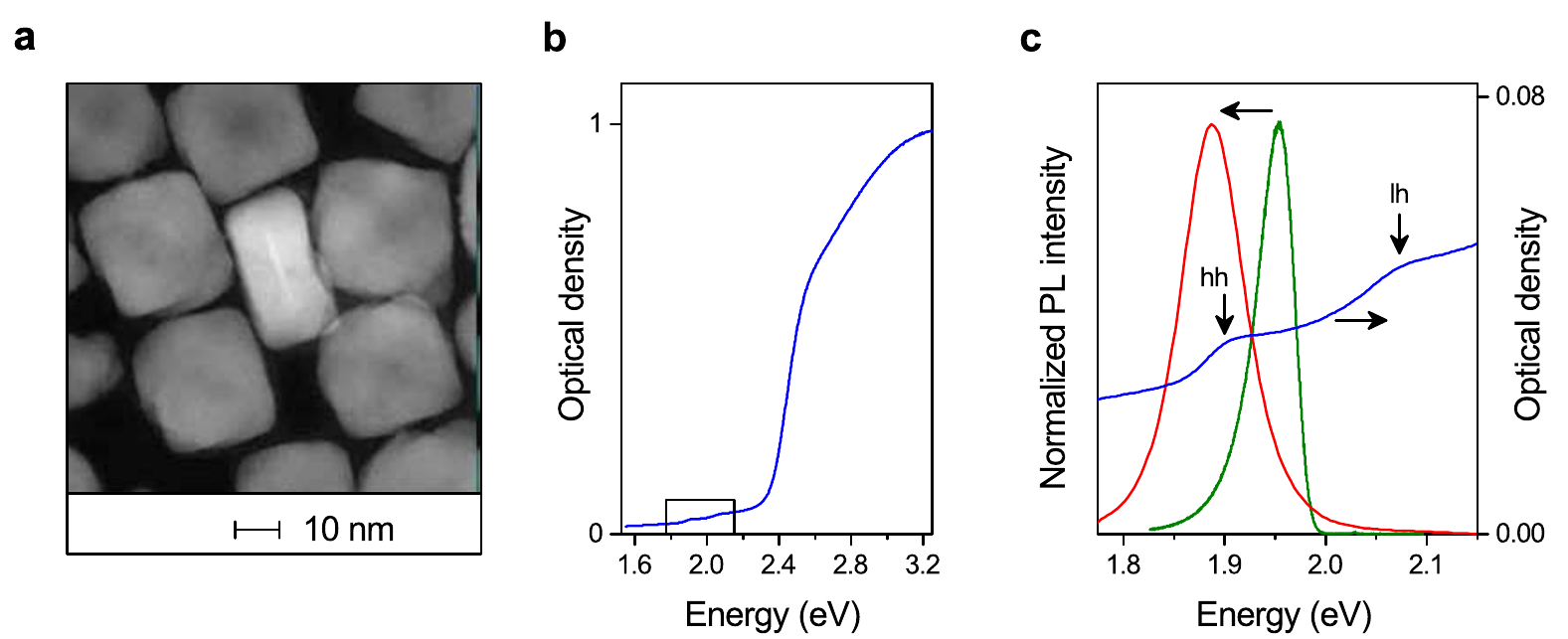}
	\caption{(a) TEM image of CdSe/CdS NPLs. Here all NPLs are lying flat and only one in the middle is staying on its narrow side. For it a brighter image of CdSe core is also seen. (b) Absorption spectrum at $T=300$~K. (c) Enlarged absorption spectrum in the region of interest at $T=300$~K (blue) and normalized PL spectra at $T=300$~K (red) and $T=4.2$~K (green).}
	\label{fig:1}
\end{figure*}

Figure~\ref{fig:1}a shows a representative transmission electron microscopy (TEM) image of the investigated CdSe/CdS NPLs. The NPLs were synthesized following modifications of a protocol published previously in Ref.~\onlinecite{Chen2013}. They have mostly square shape with $30\pm5$~nm side lengths and total thickness of $18 \pm 1$~nm contributed by 1.2 nm-thick CdSe core and $8.4 \pm 0.5$ nm-thick CdS shell. For optical measurements the samples are prepared by drop-casting concentrated NPL solutions onto a glass or silicon substrate. Optical absorption spectrum of the NPL solution shown in Figure~\ref{fig:1}b has an edge at about 2.4~eV corresponding to the band gap of CdS shell, which thickness is 14 times larger than the one of the CdSe core. CdSe absorption shown in more detail in Figure~\ref{fig:1}c (blue) has two steps at 1.900~eV and 2.074~eV corresponding to excitons formed from heavy-holes and light-holes, respectively. Their energy difference gives light-hole to heavy-hole splitting of 174~meV. Identification of the first exciton absorption peak at 1.900~eV is further confirmed by differential transmission measurements (Supporting Information, Figure~\ref{fig:Tr_Abs_Spe}).

Room temperature photoluminescence (PL) spectrum in solution has a maximum at 1.888~eV and a full width at half maximum (FWHM) of 70~meV, see Figure~\ref{fig:1}c (red), which is typical for core/shell NPLs.\cite{Ithurria2012,Tessier2013nl} Bare CdSe NPLs with thickness of 1.2~nm emit at room temperature at $\sim 2.48$~eV ($\sim 500$~nm) \cite{Biadala2014nl,Shornikova2017arXiv}. In the core-shell NPLs, similarly to core-shell spherical QDs\cite{Chen2008} and dot-in-rods,\cite{Biadala2014jp} the emission is shifted to lower energies due to spreading of the electron wave function into the shell, which reduces quantum confinement. When cooling down to $T=4.2$~K, PL shifts to 1.954~eV and narrows down to 43~meV, Figure~\ref{fig:1}c (green). Interestingly, the shape of the PL spectrum at low temperature is asymmetric evidencing that NPLs are unevenly distributed across the emission spectrum. Moreover, NPLs with higher quantum yields, \latin{i.e.} lower number of traps per NPL, typically have higher emission energies than the NPLs with higher number of traps.\cite{Tessier2013nl} This deviates the shape of the PL spectrum at low temperature from the Lorentzian or Gaussian profiles.

\section{Magneto-optical measurements.} 
\begin{figure*}[h!]
    \includegraphics{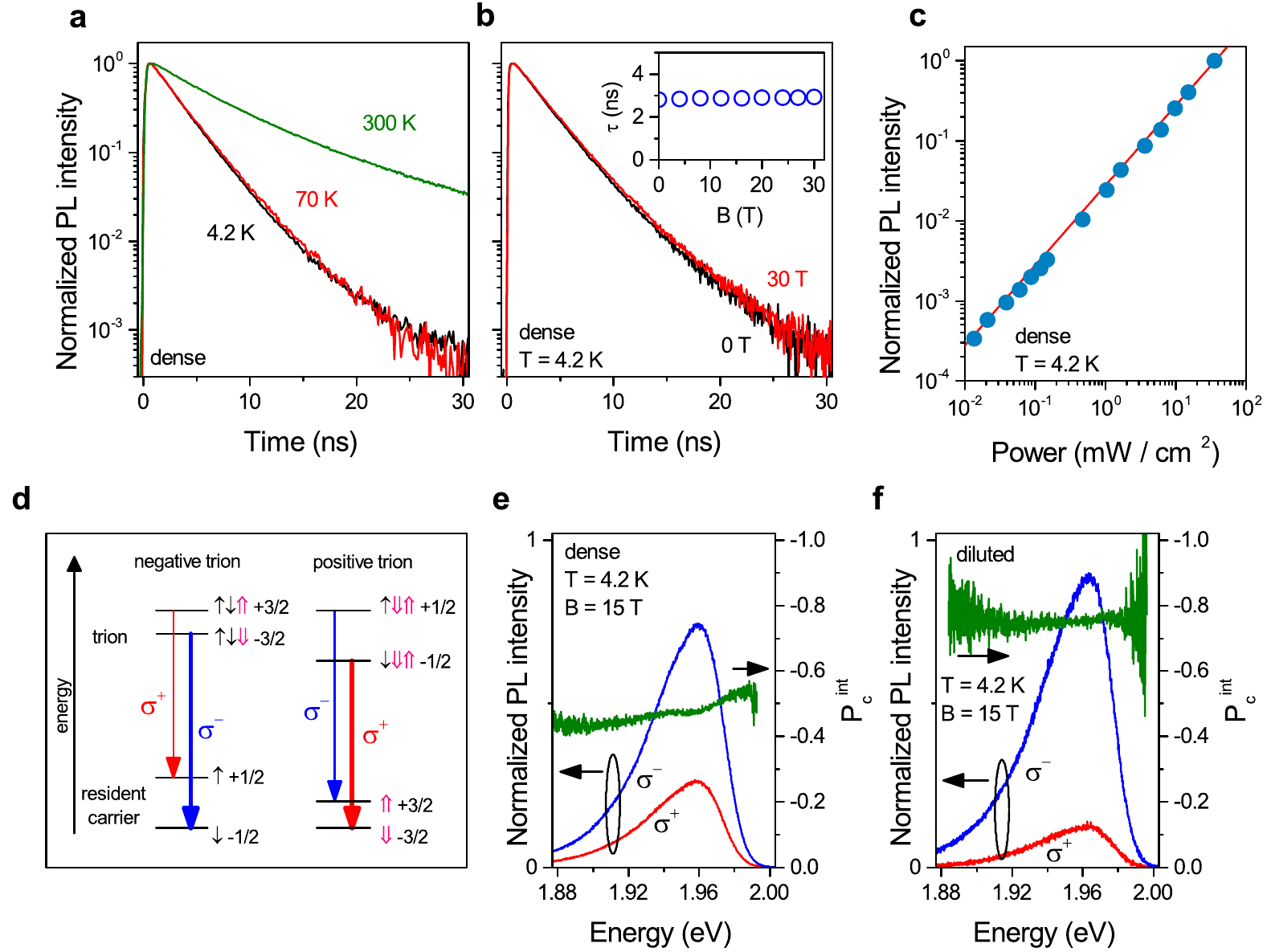}
    \caption{(a) PL decays of a dropcasted CdSe/CdS NPLs at $T=4.2$~K (black) and 70~K (red), and NPLs in solution at 300~K (green). (b) PL decays at $B=0$ (black) and 30~T (red), $T=4.2$~K. Inset: decay time dependence on magnetic field. (c) Time-integrated PL intensity dependence on excitation power. Red line is a linear fit. (d) Schematic presentation of spin structure and optical transitions for negatively and positively charged trions in magnetic field. Short black and pink arrows indicate electron and hole spins, respectively. Polarized optical transitions are shown by red ($\sigma^+$) and blue ($\sigma^-$) arrows. The more intense emission, shown by thicker arrow, comes from the lowest in energy trion state with spin $-3/2$ for the negative trion and with spin $-1/2$ for the positive trion. (e, f) Polarized PL spectra and time-integrated DCP  at $B=15$~T and $T=4.2$~K for dense and diluted samples.}
    \label{fig:TR_trion}
\end{figure*}

Recombination dynamics of a dense dropcasted sample at $T=4.2$
and 70~K are shown in Figure~\ref{fig:TR_trion}a. The PL decay is monoexponential with a decay time $\tau=3$~ns that remains constant upon increasing the temperature from 4.2 to 70~K. It is unaffected by applying magnetic field in Faraday geometry (parallel to the light beam) up to 30~T, see Figure~\ref{fig:TR_trion}b. Similar results we got for a diluted sample (Supporting Information S2). From these observations we conclude that the low temperature emission in the studied core/thick-shell NPLs is contributed not by neutral excitons, but by trions. Indeed, in an NPL without resident charge the lowest in energy  excitation of the photogenerated electron-hole pair is the optically-forbidden dark exciton with momentum projection $J_z=\pm 2$. Here the situation is the same as in colloidal CdSe QDs.\cite{Liu2013} The exciton PL dynamics at cryogenic temperatures is bi-exponential and its longer decay time related to recombination of the dark exciton is $50-100$~ns.\cite{Biadala2014nl,Shornikova2017arXiv}
The exciton decay accelerates at elevated temperatures due to population of the optically-allowed
bright state. It also becomes faster in external magnetic field tilted from the quantization axis,
which induces mixing of the dark and bright excitons.\cite{Liu2013,Shornikova2017arXiv} By contrast, the ground state of the charged exciton is optically allowed. Due to that the
time-resolved PL of a charged exciton is  monoexponential at cryogenic temperatures and
is not affected by temperature or magnetic field, as it has been reported for  CdSe/CdS thick-shell
QDs.\cite{Liu2013} From results presented in Figures~\ref{fig:TR_trion}a,b we conclude that the low
temperature emission in CdSe/CdS thick-shell NPLs is dominated by charged excitons. However, the
sign of the resident charge needs to be determined from polarization-resolved magneto-optical
experiments. Note, that the linear dependence of the emission intensity on the excitation density
(Figure~\ref{fig:TR_trion}c) excludes the biexciton contribution.

We have shown in Ref.~\onlinecite{Liu2013} that with knowledge of the signs of electron and hole
$g$-factors  the type of the resident charge carrier can be determined from the sign of PL circular
polarization induced by   external magnetic fields. In the case of a resident electron in the NPL core, a photon absorption generates a negative trion consisting of two electrons in singlet state and one heavy hole with momentum projection $M=\pm 3/2$ onto the quantization axis (low-symmetry axis perpendicular to NPLs surface). Zeeman splitting of this trion is determined by the hole $g$-factor,  $g_h$. It is strongly anisotropic and depends on NPL orientation in a magnetic field: $E_{hh}=-M g_h \mu_{\rm B} B \cos \theta$. Here $\theta \in [0;\pi]$ is the angle between the NPL quantization axis and the magnetic field direction and $\mu_{\rm B}$ is the Bohr magneton. In CdSe QDs $g_h < 0$ and the $-3/2$ level is the lowest in energy\cite{Gupta2002,Liu2013}  (left panel of Figure~\ref{fig:TR_trion}d). Here we use the definition of carrier and exciton $g$-factors and spin schemes in line with the approach of
Refs.~\cite{Efros1996,Efros2003,Liu2013} It is worthwhile to note, that for the negatively charged trion its Zeeman splitting, which determines spin polarization properties, is controlled by $g_h$, but the optical transitions from these trion states are split by the energy determined by $g_h$ and $g_e$ of the resident electron, see left panel in Figure~\ref{fig:TR_trion}d.\cite{Liu2013}

In positively charged trion the electron $g$-factor, $g_e$,  determines its Zeeman splitting (right
panel of Figure~\ref{fig:TR_trion}d). In a magnetic field the resident electron occupies one of its
Zeeman sublevels split by $E_e=S_z g_e \mu_{\rm B} B$, where $S_z=\pm 1/2$ are electron spin
projections on the magnetic field direction. The sign of $g_e$ determines which level, $-1/2$ or
$+1/2$, is the lowest in energy. In bulk CdSe, as well as in CdSe QDs, $g_e > 0$
\cite{Gupta2002,Karimov2000} and  $S_z=- 1/2$ level is the lowest.

In electric-dipole approximation, the allowed optical transitions  change the spin projection by
$+1$ or $-1$. The emitted photons are accordingly right-handed ($\sigma^+$) and left-handed
($\sigma^-$) circularly polarized. The corresponding transitions are shown by red and blue arrows
in Figure~\ref{fig:TR_trion}d. At low temperatures, when the thermal energy $kT$ ($k$ being the
Boltzmann constant) is smaller than the trion Zeeman splitting the lowest trion spin sublevel has
higher occupation and it dominates the emission.

Circularly polarized emission spectra  of the dense sample  measured at $B=15$~T are shown in
Figure~\ref{fig:TR_trion}e. The degree of time-integrated circular polarization (DCP) is defined
as: $P_c^{\rm int}=\left(I^+ - I^- \right) / \left( I^+ + I^- \right)$, where $I^+$ and $I^-$ are
time-integrated intensities of $\sigma^+$ and $\sigma^-$ circularly polarized emission, see
Supporting Information S4. It is negative in the studied CdSe/CdS NPLs. As one can see from Figure~\ref{fig:TR_trion}d, the negative DCP sign
unambiguously points toward a negatively charged trion. Therefore, we conclude that similarly to
core/thick-shell QDs\cite{Liu2013} the studied core/thick-shell NPLs are negatively charged.

As discussed above, the Zeeman spitting of the negative trion is controlled by the heavy hole,
which splitting depends on the angle $\theta$ between the NPL quantization axis and magnetic field
direction. For the NPLs horizontally oriented on the substrate, i.e.  lying flat, in the Faraday
geometry ($\theta=0^\circ$) the trion Zeeman splitting is the largest. While for the vertically
oriented NPLs, i.e. standing on the edge ($\theta=90^\circ$), no splitting is expected. This is
confirmed by comparing Figures~\ref{fig:TR_trion}e and ~\ref{fig:TR_trion}f. The latter presents
results for the NPLs from the same batch, but diluted with hexane in proportion $1:200$ before
dropcasting. This strongly changes DCP from $-0.45$ in dense NPLs up to $-0.75$ in diluted ones.
Indeed, while the dense sample contains NPLs with various orientations, in the diluted sample
almost all NPLs are horizontally oriented with $\theta \approx 0$. This is in line with experiments
on single NPL spectroscopy for different dilutions.\cite{Feng2016}

\begin{figure*}[h!]
    \includegraphics{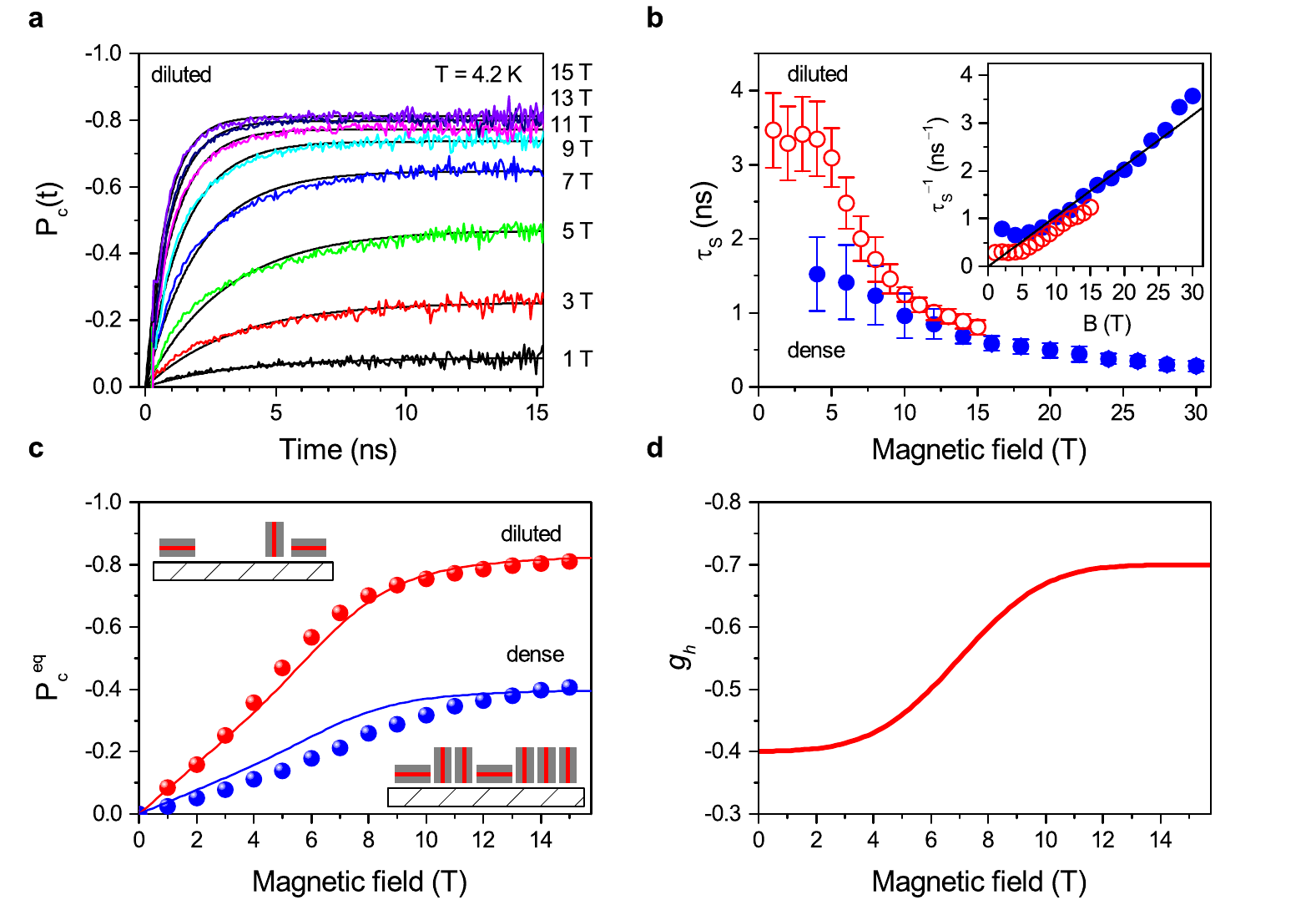}
    \caption{(a) Time-resolved DCP of diluted NPL sample measured in various magnetic fields. Lines are fits with equation \eqref{eq:Pcdyn} which allows us to evaluate  $P_c^{\rm eq}$ and $\tau_{\rm s}$.
(b) Magnetic field dependence of spin relaxation time in dense (blue) and diluted (red) ensembles.  Inset: spin relaxation rate $\tau_{\rm s}^{-1}$ dependence on magnetic field strength.  Line is linear fit with slope of 0.07~ns$^{-1}$T$^{-1}$.
(c) Magnetic field dependence of $P_c^{\rm eq}$ for dense and diluted samples. Lines are fits with equation~\eqref{eq:Psat}.
(d) Evaluated hole $g$-factor dependence on magnetic field. All data are for $T=4.2$~K.
}
   \label{fig:TR}
\end{figure*}

Figure~\ref{fig:TR}a shows time-resolved DCP: $P_c(t)=[I^+(t)-I^-(t)]/[I^+(t)+I^-(t)]$.  Just after
the laser pulse at $t=0$,  $+3/2$ and $-3/2$ spin states  of the negative trions are equally
populated and the emission is unpolarized. When the trion population relaxes to a thermal
distribution, the DCP increases reaching the equilibrium value $P_c^{\rm eq}$  at delays as long
as 15~ns in low magnetic fields. The temporal dependence of the DCP can be described by
\begin{equation}
P_c(t)=P_c^{\rm eq}(B) [1-\exp(-t/\tau_{\rm s})],
\label{eq:Pcdyn}
\end{equation}
where $\tau_{\rm s}$ is the spin relaxation time of the negative trion.  It is 3.5~ns in low magnetic fields
and shortens down to 260~ps at 30~T both for the dense and diluted NPLs, see Figure~\ref{fig:TR}b.
As one can see in the inset of Figure~\ref{fig:TR}b, the trion spin relaxation rate increases about
linearly with the field strength up to very high fields of 30~T.

Interestingly, the spin relaxation mechanism differs in thick-shell CdSe/CdS QDs and NPLs. In the former, longer spin relaxation times were measured reaching 60~ns in weak magnetic fields~\cite{Liu2013} and the spin relaxation rate was proportional to $B^2$. In NPLs the spin relaxation time is one order of magnitude faster and the spin relaxation rate increases linearly with $B$. While the latter dependence suggests an interaction with one acoustic phonon, the hole spin relaxation mechanism needs to be identified. This goes beyond the scope of this paper and further dedicated experiments are required.

Figure~\ref{fig:TR}c shows the equilibrium DCP, $P_c^{\rm eq}(B)$, measured  in the dense and
diluted samples in magnetic fields. It increases with magnetic field and is larger in the diluted
sample. Moreover, the shape of the dependences is different, i.e., the curve for the dense sample
can not be obtained by rescaling of the data  for the diluted one.

\textbf{Modeling of DCP in thick-shell NPLs.} Band offsets for conduction and valence bands  in
CdSe/CdS NPLs are  such that electrons are spreading into the CdS shell, while holes are localized
inside the CdSe core.\cite{Javaux2013} It results in large energy splitting of 174~meV between
heavy-hole and light-hole states, see Figure~\ref{fig:1}c.  As it is discussed above, the Zeeman
splitting of the negative trion is determined by the anisotropic $g$-factor of the hole confined
in the CdSe core and depends strongly on the orientation of the NPL axis and magnetic field
direction: $\Delta E_{\rm hh} = - 3g_h \mu_{\rm B}B \cos \theta$ \cite{Liu2013}.  We assume that
thick-shell CdSe/CdS NPLs tend to fall on the surface of substrate predominately horizontally
($\cos \theta =1$)  or vertically ($\cos \theta =0$). This assumption is in agreement with TEM
analysis of our samples (Figure~\ref{fig:1}a) and results reported in Ref.~\onlinecite{Feng2016}.  In
Faraday geometry, the magnetic field  directed perpendicular to a substrate with drop-casted NPLs
induces the circular polarization of PL only in horizontally aligned NPLs. The PL in the vertically
aligned NPLs  is linearly polarized in the NPL plane. The contribution from the vertically aligned
NPLs  to the total intensity of the PL signal decreases the  maximum achievable value of DCP in
strong magnetic fields $|P_c^{\rm sat}| = |P_c^{\rm eq}(B \to \infty)|$  according to
$|P_c^{\rm sat}|=2n_{\rm hor}/(2n_{\rm hor}+n_{\rm vert})$, where $n_{\rm hor}$ and $n_{\rm vert}$
are the fractions of NPLs lying horizontally or vertically, respectively (Supporting Information
S5). Taking into account that $n_{\rm hor}+n_{\rm vert}=1$, we find the relation between $|P_c^{\rm
sat}|$ and $n_{\rm hor}$, that is $|P_c^{\rm sat}|=2n_{\rm hor}/(n_{\rm hor}+1)$.

To evaluate $g_h$ and $n_{\rm hor}$ we analyze  the value of the equilibrium DCP, $P_c^{\rm eq}(t
\to \infty)$,  measured under pulsed excitation. Its magnetic field dependence can be described by
the following equation (Supporting Information S5):
\begin{equation}
P_c^{\rm eq}(B)= |P_c^{\rm
sat}|  \tanh \frac{3g_{h}\mu_{\rm B} B}{2kT} .
\label{eq:Psat}
\end{equation}
Using the relation between $|P_c^{\rm sat}|$ and $n_{\rm hor}$ we find from the experimental values of $|P_c^{\rm sat}| = 0.83$ and $0.40$   the fractions of horizontally aligned NPLs $n_{\rm hor} = 0.70$ in diluted and 0.25  in dense ensembles (Figure~\ref{Fig4}). It indicates that high concentration of NPLs in dense solution results in their interaction, leading to their preferable vertical orientation in the dropcasted sample. By contrast, non-interacting single NPLs in diluted solution tend to settle on substrate horizontally. Note, that in ensemble of spherical colloidal QDs with random orientations the DCP saturation should be at $-0.75$. Same is valid for randomly oriented NPL ensemble (Figure~\ref{Fig2}). Therefore, in the studied sample the presence of two preferable orientations is confirmed by the fact, that both in the dense and diluted samples $|P_c^{\rm sat}| \neq 0.75$.

Knowing the value of $|P_c^{\rm sat}|$ we determine $g_{ h}$  from fitting the experimental
magnetic field dependences of $P_c^{\rm eq}(B)$. In weak magnetic fields below 4~T,  $P_c^{\rm
eq}(B)$ increases linearly with magnetic field with $g_h=-0.40$. This value is in reasonable
agreement with the theoretically calculated  value of $g_{ h}$ for CdSe quantum well  (see below
and Supporting Information S6). However, the DCP dependence in the whole field range can not be
fitted with the constant $g_h=-0.40$  (Figure~\ref{Fig3}). The best fit
shown in Figure~\ref{fig:TR}c required the use of the magnetic field dependent $g_h(B)$ shown in
Figure~\ref{fig:TR}d. One can see that in high magnetic fields $g_h$ approaches $-0.70$. This
value is close to the $g$-factor in spherical QDs: for thick-shell CdSe/CdS QDs
$g_h=-0.54$,\cite{Liu2013} for bare core CdSe QDs $g_h=-0.73$.\cite{Gupta2002} As we discuss
below and in Supporting Information S6, the value of the hole $g$-factor depends strongly on the
mixing of light-hole and heavy-hole states and on the spatial distribution of the hole charge
density (wave functions) and thus on the type of confining potential.

\section{Spin-flip Raman scattering.} 
\begin{figure*}[h!]
    \includegraphics{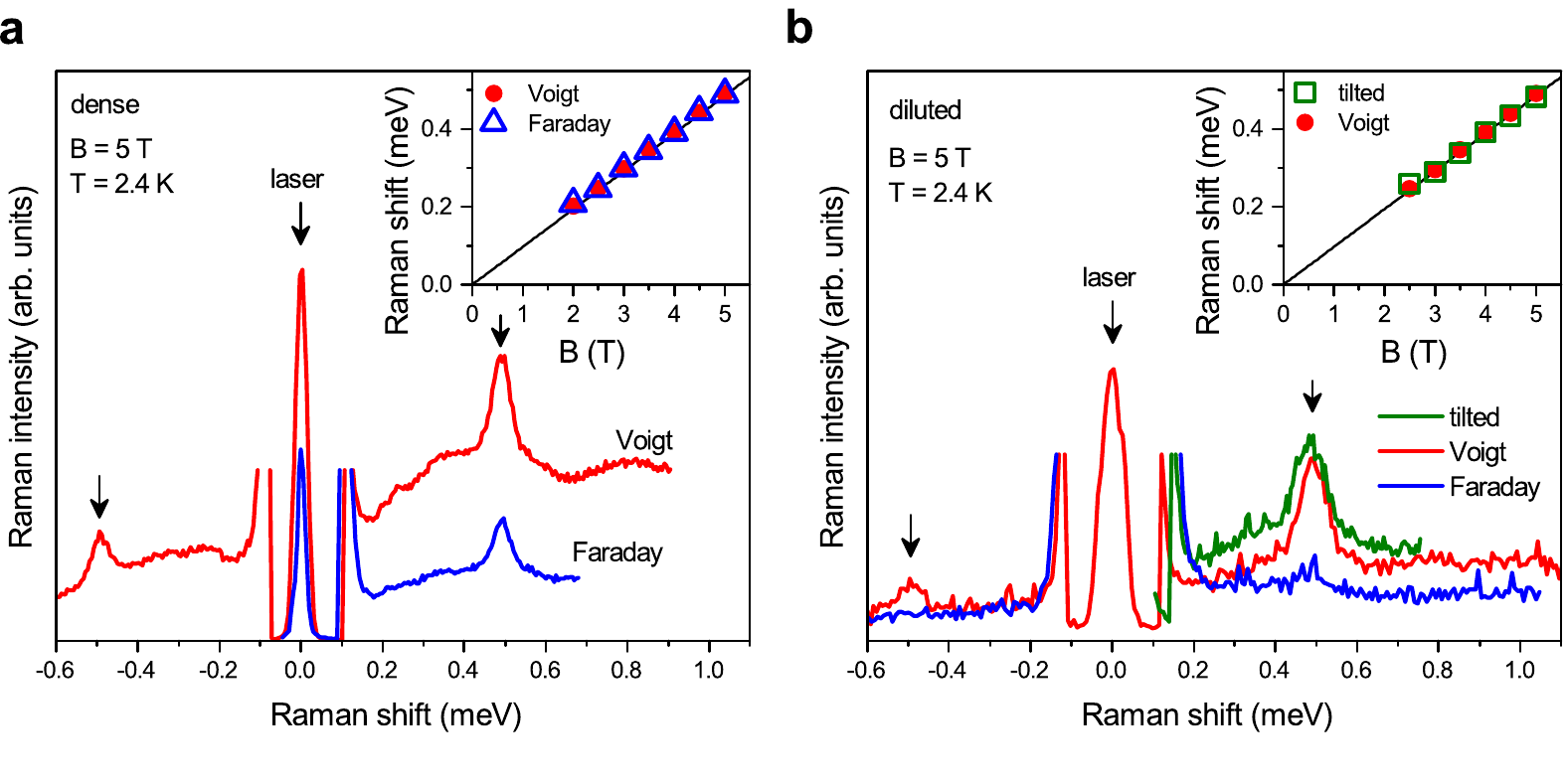}
    \caption{Spin-flip Raman scattering spectra measured at $B=5$~T for NPLs dropcasted onto a silicon substrate: (a) dense and (b) diluted samples. Electron spin-flip lines are marked by arrows. Geometries are Voigt (red), Faraday (blue) and tilted by $20^{\circ}$ from Faraday (green). Crossed linear polarizations are used for Voigt geometry and copolarized circular polarizations for Faraday and tilted geometries. Laser photon energy is 1.9588~eV. Stokes signals correspond to positive Raman shifts. Insets: magnetic field dependence of the Raman shifts of the electron spin-flip lines. Lines are linear fits.}
    \label{fig:SFRS}
\end{figure*}

We measure the electron $g$-factor by means of the spin-flip
Raman scattering technique.  The energy shift of the spin-flip Raman scattering (SFRS) line is directly related to Zeeman
splitting of the resident electron and allows to evaluate
$g_e$.\cite{Sirenko1997,Debus2013,Debus2014} Figure~\ref{fig:SFRS}a shows SFRS spectra measured on
the dense sample in Voigt (magnetic field perpendicular to the light beam) and Faraday (magnetic
field parallel to the light beam) geometries at $B = 5$~T and $T = 2.4$~K. The zero of the horizontal
axis corresponds to the laser energy. The SFRS lines are shifted from the laser energy by $+0.49$
and $-0.49$ meV for the Stokes and anti-Stokes lines, respectively. Background signal in spectra is due
to the resonant PL. In both geometries the Raman shifts have same values (inset
Figure~\ref{fig:SFRS}a). They depend linearly on magnetic field. It allows us to evaluate $g_e =
1.68 \pm 0.01$ for the dense sample and $g_e = 1.69 \pm 0.02$ for the diluted one.
Keeping in mind that the dense and diluted samples have different ratio of horizontally and vertically oriented NPLs, their close $g_e$ values allows us to conclude on the high isotropy of the electron
$g$-factor in the studied CdSe/CdS NPLs.

It is instructive to compare polarization properties of SFRS  lines for the dense and diluted
samples, as the Raman optical selection rules are sensitive to the NPL orientation and, therefore, can provide information on it. In the dense sample the lines have equal intensities in co- ($\sigma^+ / \sigma^+$) and cross- ($\sigma^+ / \sigma^-$)  circular polarizations in Faraday geometry and as wells as in parallel (V/V) and crossed (H/V) linear polarizations in Voigt geometry (not shown). Such behavior is in agreement with our conclusion made from DCP data on the presence of NPLs with horizontal and
vertical orientations in the dense ensemble.  In contrast, in the diluted sample with predominant
horizontally oriented NPLs the SFRS line is very weak in Faraday geometry (Figure~\ref{fig:SFRS}b,
blue) and becomes strong when magnetic field is tilted by small angle (Figure~\ref{fig:SFRS}b,
green). In Voigt geometry the SFRS line is more pronounced in crossed linear polarizations
(H/V) (Figure~\ref{fig:SFRS}b, red). The very same properties of the SFRS lines of resident electrons have been reported for (In,Ga)As/GaAs singly-charged epitaxial QDs.\cite{Debus2014} These dots have defined orientation with the heavy-hole quantization axis along the structure growth axis. The similarity in experimental appearances confirms the dominant horizontal orientation of NPLs in the diluted sample. In the dense sample, presence of horizontal and vertical NPLs results in effective violation of
the strict selection rules for SFRS.

\section{Pump-probe Faraday rotation.}
\begin{figure*}[h!]
	\includegraphics{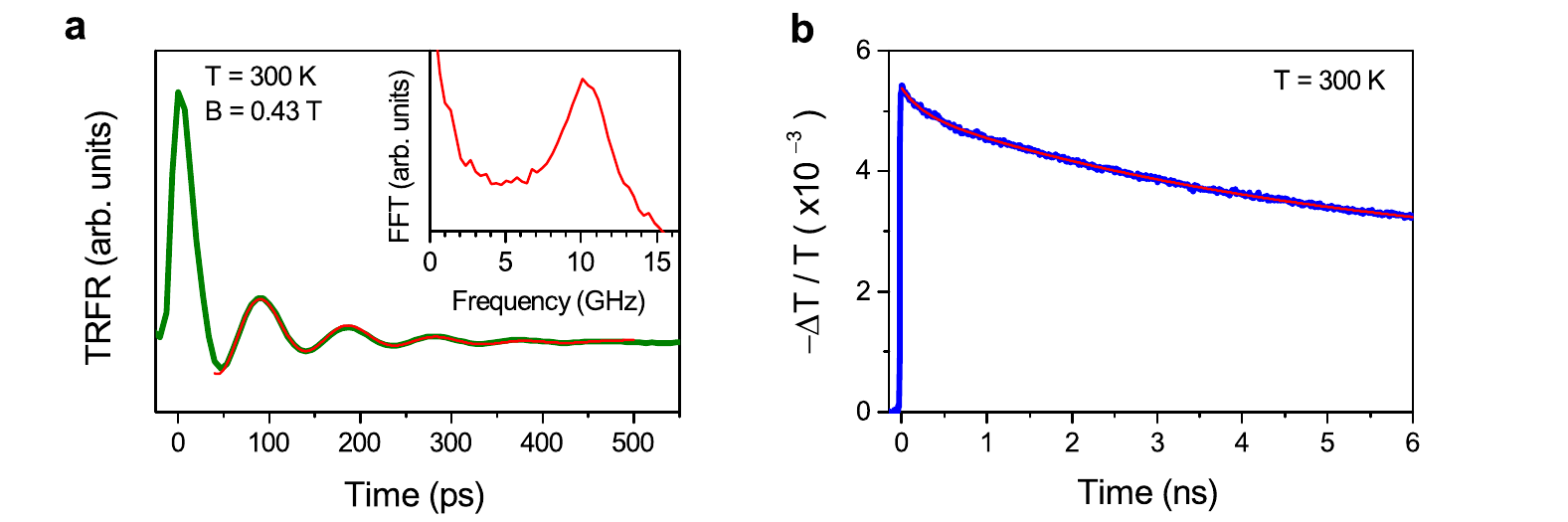}
	\caption{(a) Time-resolved Faraday rotation measured for NPLs in solution (green) at 1.900~eV together with fit by equation~\eqref{eq:quantum_beats} (red) with parameters $T_2^*=100$~ps and $|g_e|=1.71 \pm0.03$. Inset: FFT spectrum reveals single precession frequency.
		(b) Time-resolved differential transmission at 1.900~eV (first exciton absorption peak). Red line is bi-exponential fit with times of 240~ps and 5~ns.
	}
	\label{fig:pump-probe}
\end{figure*}

To measure the electron $g$-factor at room temperature in the NPL solution we use a time-resolved pump-probe Faraday rotation (TRFR) technique,\cite{Chapter6} which has been successfully exploited for investigating coherent spin dynamics in colloidal QDs.\cite{Gupta1999,Gupta2002,Li2014,Masumoto2015,Feng2017} By contrast with DCP, which addresses the spin dynamics of photogenerated trions, the TRFR allows to address spin coherence of photoexcited and  resident carriers. We used a pulsed laser system based on a regenerative laser amplifier combined with a
narrow band optical parametric amplifier (OPA) generating pulses with duration of 3~ps (Supporting Information S1). The OPA photon energy was tunable in a wide spectral range, and for this experiment it was set at 1.900~eV which corresponds to the exciton absorption. The laser beam was split into pump and probe beams. The circularly polarized pump pulses generate spin polarization, whose dynamics were monitored via the rotation of polarization plane of the linearly polarized probe pulses by tuning the pump-probe time delay. Solution with CdSe/CdS NPLs was in quartz cuvette placed between a permanent magnet pair  providing a transverse magnetic field of  $B=0.43$~T.
The generated spin polarization precesses around the field direction  with a Larmor frequency $\omega_{\rm L}=g_e \mu_{\rm B}B/\hbar$, which was detected as time-resolved spin beats shown in Figure~\ref{fig:pump-probe}a. The spin signal decays with the ensemble spin dephasing time $T_2^*$  due to the various  mechanisms: electron-hole recombination, electron-nuclear hyperfine interaction, and inhomogeneous dephasing.\cite{Li2014}
The red curve shows the fit with an exponentially damped oscillation:
\begin{equation}
I_{\rm FR}(t)=I_{\rm FR}(t=0)\exp(-t/T_2^*)\cos(\omega_{\rm L} t),
\label{eq:quantum_beats}
\end{equation}
The presence of only one oscillation frequency corresponding to $|g_e|=1.71 \pm0.03$  is confirmed by the fast Fourier transform (FFT) spectrum  in the inset of Figure~\ref{fig:pump-probe}a. The spin polarization  looses its coherence with time $T_2^*=100$~ps, which is faster than that in CdSe colloidal QDs falling in range from 400~ps to 3~ns.\cite{Gupta1999,Fumani2013}

Hole spin relaxation at room temperature typically is in a subpicosecond range due to strong spin-orbit interaction in semiconductor nanostructures. The exciton spin relaxation dynamics is also fast in colloidal CdSe QDs. At room temperature it covers the range from 200~fs to 2~ps depending on the QD size and cubic or wurtzite crystal structure.\cite{Huxter2010} We observed spin beats in the Faraday rotation signal at much longer time delays of 100~ps. Therefore, we conclude that this signal originates from either resident electrons in negatively charged NPLs or photogenerated electrons in positively charged trions, i.e. the NPLs can be positively charged at room temperature.\cite{Feng2017}  Note, that from these experiments we cannot say what fraction of NPLs in the excited ensemble is charged and what stays neutral.

Differential transmission dynamics shown in Figure~\ref{fig:pump-probe}b reflects the population dynamics of NPLs with photogenerated carriers. At room temperature it can be fitted by bi-exponential decay with times of 240~ps and 5~ns. The fast component belongs to recombination of multi-exciton complexes, typically limited by fast Auger recombination.\cite{Li2017}  The longer process provides the main part of the signal. Its decay time is in agreement with PL recombination decay time of 5~ns at room temperature given in Figure~\ref{fig:TR_trion}a.


\section{Calculation of electron $g$-factor.} Isotropic electron $g$-factor and its weak temperature
dependence  can be explained by the fact that the electron charge density is mostly localized in
the CdS shell. Indeed, if one neglects the conduction band offset at the CdSe/CdS interface, one
can evaluate the electron $g$-factor in core-shell NPL as $g_e=g_e^{\rm CdSe} w^{\rm CdSe} +
g_e^{\rm CdS} (1- w^{\rm CdSe})$, where $g_e^{\rm CdSe} = 0.42$\cite{Karimov2000}  for cubic c-CdSe
and  $g_e^{\rm CdS} = 1.78$\cite{Hopfield1961,Sirenko1998}  for CdS are bulk values and $w^{\rm CdSe}$ is the probability to find the electron in the CdSe core. The experimental value of $g_{\rm e}=1.68$
allows us to estimate $w^{\rm CdSe} \approx 0.17$.

\section{Calculation of hole $g$-factor.}
\begin{table*}
	\begin{center}
		\caption{Hole $g$-factors calculated for cubic and wurtzite CdSe-based nanostructures. Results for other II-VI semiconductor nanostructures are given in Table~\ref{Table_g2}.}
		\label{Table_g1}
		\begin{tabular}{| l | l | l | l |  l | l | l |  l | l |  l | l |}
			\hline
			Material & $\gamma_1$ & $\gamma$ & $\varkappa$&$g_{\rm bulk}$ &$\beta$&$g_{\text{sph}}^{(\text{par})}$&$g_{\text{sph}}^{(\text{box})}$&
			$g_{\text{QW}}^{(\text{par})}$&$g_{\text{QW}}^{(\text{box})}$&Refs.\\ \hline
			c-CdSe &5.51 & 1.78 &0.4633&0.926 &0.215&-0.78&-0.94&-0.113&0.36&\onlinecite{Adachi}  \\ \hline
			c-CdSe & 3.265 & 1.33 & 0.4627 &0.925 &0.1&-0.83&-1.12&-0.13&0.46&\onlinecite{Karazhanov}\\ \hline
			w-CdSe& 2.52 & 0.83 & -0.1233&-0.246 &0.2&-0.98&-1.05&-0.74&-0.5&\onlinecite{Fu} \\ \hline
			w-CdSe& 1.7 & 0.4 & -0.567&-1.133 &0.36&-1.32&-1.29&-1.3&-1.23&\onlinecite{Kapustina} \\ \hline
			w-CdSe& 2.04 & 0.58 &-0.38&-0.76 &0.275&-1.16&-1.12&-1.05&-0.93&\onlinecite{Norris} \\ \hline
			w-CdSe& 2.1 & 0.55 &-0.45&-0.9 &0.312&-1.23&-1.19&-1.16& -1.05&\onlinecite{Ekimov} \\ \hline
			w-CdSe& 1.67 &0.56 &-0.29&-0.58 &0.197&-1.04&-1.09&-0.92&-0.76&\onlinecite{Berezovsky} \\ \hline
		\end{tabular}
	\end{center}
\end{table*}

As the holes are confined mostly in the CdSe core we consider the four time degenerate $\Gamma_8$
valence subband for semiconductors with  large spin-orbit splitting and neglect the spin-orbit split valence subband. In this case the hole Hamiltonian can be written  as
\begin{equation}\label{Hamilt}
\widehat{H}=\widehat{H}_L+\widehat{H}_{\text{Z}}+\widehat{H}_B+V_{\rm ext}({\bf r}).
\end{equation}
Here
\begin{equation}\label{lutt}
\widehat{H}_L=\frac{\hbar^2}{2 m_0}\left[\left(\gamma_1+\frac{5}{2}\gamma\right)k^2-2\gamma({\bf kJ})^2\right],
\end{equation}
is the Luttinger Hamiltonian in spherical approximation in zero magnetic field\cite{Luttinger,Gelmont71}. We neglect the  valence band warping in cubic semiconductors as the
spherical approximation is sufficient for estimation of the hole $g$-factor.  ${\bf J}$ is
the hole internal angular momentum operator for $J=3/2$, $m_0$ is the free electron mass, and
$\gamma_1$ and $\gamma=(2\gamma_2+3\gamma_3)/5$ are Luttinger parameters related to the light-hole
and heavy-hole effective masses as $m_{\rm l,h} = m_0/(\gamma_1 \pm 2 \gamma)$. $V_{\rm ext}({\bf
r})$ is the quantum structure potential determining the spatial distribution of the hole density.
The magnetic field contributions $\widehat{H}_{\text{Z}}$ and $\widehat{H}_B$ are considered in
detail in Supporting Information S6.

The value of the hole effective  $g$-factor depends on the hole wave function and, therefore, on
the potential acting on the hole. For several potential shapes it is possible to make simple
estimations for the hole $g$-factor. Here we consider spherical and quantum well-like potential of
two types: parabolic potential and box-like potential   with abrupt infinite barriers. It can be
shown, that for Gaussian-like smooth potential with the finite barriers hole effective $g$-factor
in most cases is the same as for the parabolic potential for confined ground state\cite{Semina2016}. In the case of the quantum
well-like potential the external magnetic field is assumed to be directed along the structure
growth axis.  We find that the values of the hole $g$-factors are determined by  symmetry and type
of $V_{\rm ext}({\bf r})$ potential.

Calculated hole $g$-factors for cubic and wurtzite CdSe-based structures are given in
Table~\ref{Table_g1}. The sets of $\gamma_1$ and $\gamma$ Luttinger parameters were taken from
literature, magnetic Luttinger parameter $\varkappa$ was evaluated by approximate formula
\cite{Roth}:
\begin{equation}\label{kappa}
\varkappa\approx-2/3+5\gamma/3-\gamma_1/3 .
\end{equation}
For the bulk CdSe the heavy hole $g$-factor $g_h = 2\varkappa$ is given in $g_{\rm bulk}$ column.
The external potential $V_{\rm ext}({\bf r})$ mixes the light-hole and heavy-hole Bloch states
depending of the potential type and symmetry and the value of the light-hole to heavy-hole
effective mass ratio $\beta=(\gamma_1-2\gamma)/(\gamma_1+2\gamma)$. The columns
$g_{\text{sph}}^{(\text{par})}$ and $g_{\text{sph}}^{(\text{box})}$ show  the hole ground state
effective $g$-factors $g_h$ in spherical quantum dots  calculated with parabolic and box-like
potential, respectively, while $g_{\text{QW}}^{(\text{par})}$ and $g_{\text{QW}}^{(\text{box})}$
correspond to $g_h$ in thin QW. By analogy, we assume that $g_h$ in NPL are also given by
$g_{\text{QW}}^{(\text{par})}$ and $g_{\text{QW}}^{(\text{box})}$. Note, that the
estimated values of hole $g$-factors in all configurations are independent of the characteristic size of the the localization potential,  for details of calculations see Supporting Information S6. As one
can see, in spherical quantum dots $g_h$ are in the range from $-0.8$ to $-1.2$ and are weakly
dependent on the potential type and crystal structure. By contrast, for quantum well-like
structures the $g$-factor for cubic CdSe (c-CdSe) depends on the potential type, and, importantly,
on the parametrization: it can have different sign, for example.  For wurtzite CdSe (w-CdSe)
$g$-factors fall almost in the same ranges as for spherical crystals for both potential types and
all used parametrizations.

In the studied CdSe/CdS NPLs, the $g_h$ values determined experimentally fall in the range from
$-0.4$ to $-0.7$, that is in between  estimations for quantum well-like structures with the
parabolic-like potential and the spherical quantum dots for c-CdSe.  We note, that although the
spatial structure potential for the holes in the direction of the NPL anisotropic axis is the
abrupt one, it can be smoothed by the Coulomb potential of two electrons forming the trion.
In addition, the increase of magnetic field  seems to affect the spatial distribution
of the hole density causing the field dependence of $g_h(B)$ (Figure~\ref{fig:TR}d). In this case  one
has to consider the magnetic field dependent potential $\hat H_{\rm B}$ (see equation (S11) in
Supporting Information S6) as the part of the external confining potential directly and not as the
perturbation. This magnetic-field-induced potential may considerably modify the  mixing between
heavy-hole and light-hole subbands  even in the magnetic field parallel to the quantization axis.
Such calculations, however, are beyond the scope of the present paper.

In summary, comprehensive magneto-optical studies involving several time-resolved and polarization-resolved techniques in high magnetic fields have allowed unveiling the origin of the emission in thick shell CdSe/CdS nanoplatelets. We show that it is compelled by the radiative recombination of the negative trions. Thorough analysis leads to evaluation of the electron and hole $g$-factors and the spin relaxation dynamics of the negatively charged trions. Interestingly, we show that the orientation of the nanoplatelets in the dropcasted samples depends strongly on the dilution of the initial solution and develop a magneto-optical technique for the quantitative  evaluation of the nanoplatelets orientation in ensemble measurements. These studies highlight that the experimental approaches well-settled for investigation of spin-dependent phenomena in epitaxially grown nanostructures are very prospective for new colloidal nanomaterials.


\section{ASSOCIATED CONTENT}

\textbf{Supporting Information.}
Details of experimental techniques, statistics for DPC in samples with various dilutions, theoretical approach to DCP of negatively charged trions, modeling of DCP in thick-shell NPLs, calculation of hole $g$-factor in II-VI semiconductor nanostructures.

\section{ACKNOWLEDGEMENTS}

The authors are thankful to  Al.L. Efros for fruitful discussions.
E.V.S., D.H.F., D.R.Y., A.V.R., M.A.S., A.A.G., and M.B. acknowledge support of the Deutsche Forschungsgemeinschaft in the frame of ICRC TRR 160. E.V.S., D.R.Y., V.F.S. and Yu.G.K. acknowledge the Russian Science Foundation (Grant No. 14-42-00015). We acknowledge the support from HFML-RU/FOM, a member of the European Magnetic Field Laboratory (EMFL). B.D. acknowledges funding from the EU Marie Curie project 642656 "Phonsi".

\onecolumn
\newpage

\begin{center}
	\section{Supporting information}
	
	\textbf{\large Electron and hole $g$-factors and spin dynamics of negatively charged excitons in CdSe/CdS colloidal nanoplatelets with thick shells}
	
	E. V. Shornikova$^{1,2,*}$, L. Biadala$^{3,*}$, D. R. Yakovlev$^{1,4,*}$, D. H. Feng$^1,5$, V. F. Sapega$^{4}$, N. Flipo$^1$,  A. A. Golovatenko$^{4}$, M. A. Semina$^{4}$, A. V. Rodina$^{4}$, A. A. Mitioglu$^6$, M. V. Ballottin$^6$, P. C. M. Christianen$^6$, Yu. G. Kusrayev$^{4}$, M. Nasilowski$^7$, B. Dubertret$^7$, and M. Bayer $^{1,4}$
	
\end{center}

$^1$\textit{Experimentelle Physik 2, Technische Universit\"{a}t Dortmund, 44227 Dortmund, Germany}

$^2$\textit{Rzhanov Institute of Semiconductor Physics, Siberian Branch of Russian Academy of Sciences, 630090 Novosibirsk, Russia}

$^3$\textit{Institut d'Electronique, de Micro{\'e}lectronique et de Nanotechnologie, CNRS, 59652 Villeneuve-d'Ascq, France}

$^4$\textit{Ioffe  Institute, Russian Academy of Sciences, 194021 St. Petersburg, Russia}

$^5$\textit{State Key Laboratory of Precision Spectroscopy, East China Normal University, Shanghai 200062, China}

$^6$\textit{High Field Magnet Laboratory (HFML-EMFL),Radboud University, 6525 ED Nijmegen, The Netherlands}

$^7$\textit{Laboratoire de Physique et d'Etude des Mat\'{e}riaux, ESPCI, CNRS, 75231 Paris, France}

$^*$Corresponding Authors:

elena.kozhemyakina@tu-dortmund.de; 
louis.biadala@isen.iemn.univ-lille1.fr;
dmitri.yakovlev@tu-dortmund.de

\bigskip

\setcounter{equation}{0}
\setcounter{figure}{0}
\setcounter{table}{0}
\setcounter{page}{1}
\renewcommand{\theequation}{S\arabic{equation}}
\renewcommand{\thefigure}{S\arabic{figure}}
\renewcommand{\thepage}{S\arabic{page}}
\renewcommand{\thetable}{S\arabic{table}}

\subsection{S1. Experimental techniques}

\textbf{Recombination and spin dynamics measurements} at low temperatures of $T=4.2$ and 70~K  were performed on the dropcasted on a glass plate samples being in contact with helium exchange gas. Room temperature experiments were made for NPLs in solution. External magnetic fields up to 15~T, generated by a superconducting solenoid, were applied in the Faraday geometry, i.e. parallel to the optical axis.
Photoluminescence (PL) was excited with a pulsed diode laser (photon energy 3.06~eV (405~nm), pulse duration 50~ps, repetition rate between 0.8 and 5~MHz) with a weak average excitation power density ($<0.02$~W/cm$^2$).
The PL in backscattering geometry passed through a combination of a quarter-wave plate and a linear polarizer. By rotating the quarter-wave plate, the two circularly polarized components of the PL, $\sigma^+$ and $\sigma^-$ were measured independently.
The PL signal was dispersed by a 0.55-m spectrometer and detected by a liquid-nitrogen-cooled charge-coupled-device (CCD) or an avalanche Si-photodiode (APD) connected to a conventional time-correlated single-photon counting setup. The temporal resolution of our setup is 100~ps.

\textbf{Experiments in magnetic fields up to 30~T}. These measurements were performed in High Field Magnet Laboratory, Nijmegen. The NPL sample was mounted in a titanium sample holder on a top of a three-axis piezo-positioner. The sample stage was placed in an optical probe, made of carbon and titanium to minimize possible displacements at high magnetic fields. Laser light was focused on the sample by a lens (10~mm focal length). The same lens was used to collect the PL emission and direct it to the detection setup (backscattering geometry). The optical probe was mounted inside a liquid helium bath cryostat (4.2~K) inserted in a 50~mm bore Florida-Bitter electromagnet with a maximum $dc$ magnetic field strength of 31~T. Experiments were performed in Faraday geometry (light excitation and detection parallel to the magnetic field direction). For time-resolved PL measurements the excitation was provided by a picosecond pulsed diode-laser operating at 405~nm (photon energy 3.06~eV). The PL signal was detected by an avalanche Si-photodiode connected to a single-photon counter (conventional time-correlated single-photon counting setup).

\textbf{Spin-flip Raman scattering (SFRS).}
For the excitation of SFRS, we used the line of He-Ne laser (632.8~nm, 1.9588~eV). Laser power densities focused on the sample did not exceed 10~Wcm$^{-2}$. Scattered light was analyzed by a Jobin-Yvon U1000 double monochromator equipped with a cooled GaAs photomultiplier and conventional photon counting electronics. The spectral slits width of 0.2~cm$^{-1}$ (0.024~meV)  and 0.5~cm$^{-1}$ (0.06~meV)  were set for measurements of the dense and diluted NPLs, respectively. Raman spectra were measured in crossed linear polarizations for  Voigt geometry and in copolarized circular polarizations for Faraday and tilted  geometries.

\textbf{Time-resolved differential transmission and pump-probe Faraday rotation \newline (TRFR).}
The time-resolved pump-probe setup is based on a regenerative amplifier Yb-KGW (Ytterbium doped potassium gadolinium tungstate) laser system (PHAROS, Light Conversion Ltd.) combined with a narrow band picosecond optical parametric oscillator (ps-OPA) and a broadband fs-OPA (ORPHEUS-PS and ORPHEUS, respectively, Light Conversion Ltd.), which are operated at a repetition frequency of 30~kHz.  At the wavelengths used in this study, the pulse duration is 3~ps for ps-OPA and 180~fs for fs-OPA. The laser linewidth is 0.55~nm for ps-OPA and 6.3~nm for fs-OPA.

For the pump-probe Faraday rotation experiment the laser beam was split from ps-OPA for degenerate pump and probe beams. The delay time between pump and probe pulses was scanned by means of a mechanical delay line. Experiments were performed at room temperature with sample solution in quartz cuvette. The pump fluence was typically 50~$\mu$J/cm$^2$ and the probe one about 10 times smaller. Circular polarization of the pump beam was modulated between $\sigma^+$ and $\sigma^-$ by an electro-optical modulator. Probe beam was linearly polarized. The linear polarization plane of the probe pulse transmitted through the sample was rotated due to Faraday effect. The Faraday rotation angle was detected by a polarization-sensitive Wollaston beam splitter and a sensitive balanced photodiode interfaced by a lock-in amplifier. Experiment was performed in Voigt geometry with an external magnetic field applied perpendicular to the light wave vector. For that the cuvette with NPL solution was placed between the poles of a permanent magnet pair generating field of $B=0.43$~T.

\begin{figure*}[h!]
    \includegraphics[width=8cm]{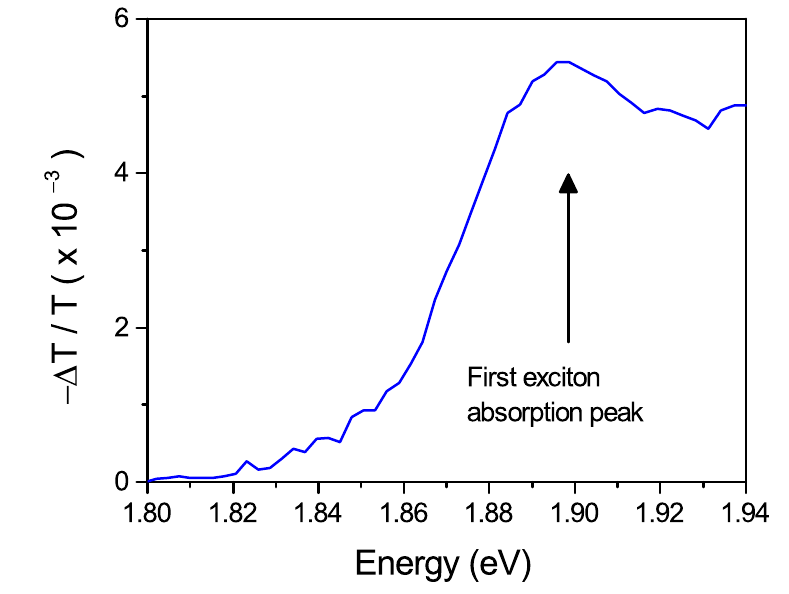}
    \caption{Differential transmission spectrum of CdSe/CdS NPLs in solution. $T=300$~K.}
    \label{fig:Tr_Abs_Spe}
\end{figure*}

The time-resolved differential transmission measurements are either performed by degenerate pump-probe from ps-OPA (Figure~\ref{fig:pump-probe}b in the main text, with the same laser conditions as in Figure~\ref{fig:pump-probe}a for Faraday rotation measurements) or by fs-pump and ps-probe for spectral dependence of the differential transmission (Figure~\ref{fig:Tr_Abs_Spe}). Figure~\ref{fig:Tr_Abs_Spe} was measured by scanning the probe photon energy in the spectral range from 1.80 to 1.94~eV, with fixed fs-pump energy at 1.90~eV. The peak at 1.900~eV in the $-\Delta T/T$ spectrum corresponds to the exciton absorption. Note that we are working in the linear regime, where the differential transmission is proportional to the population of photogenerated carriers and/or excitons.

\subsection{S2. Recombination dynamics in diluted sample}

Recombination dynamics of a diluted dropcasted sample are shown in Figure~\ref{fig:SI_TR(T,B_diluted)}. Similar to the  dense sample, the PL decay is monoexponential with a decay time $\tau=3$~ns that remains constant upon increasing temperature from $T=4.2$ to $70$~K (Figure~\ref{fig:SI_TR(T,B_diluted)}a), and  magnetic field from $B=0$ to $15$~T (Figure~\ref{fig:SI_TR(T,B_diluted)}b).

\begin{figure*}[h!]
    \includegraphics[width=14cm]{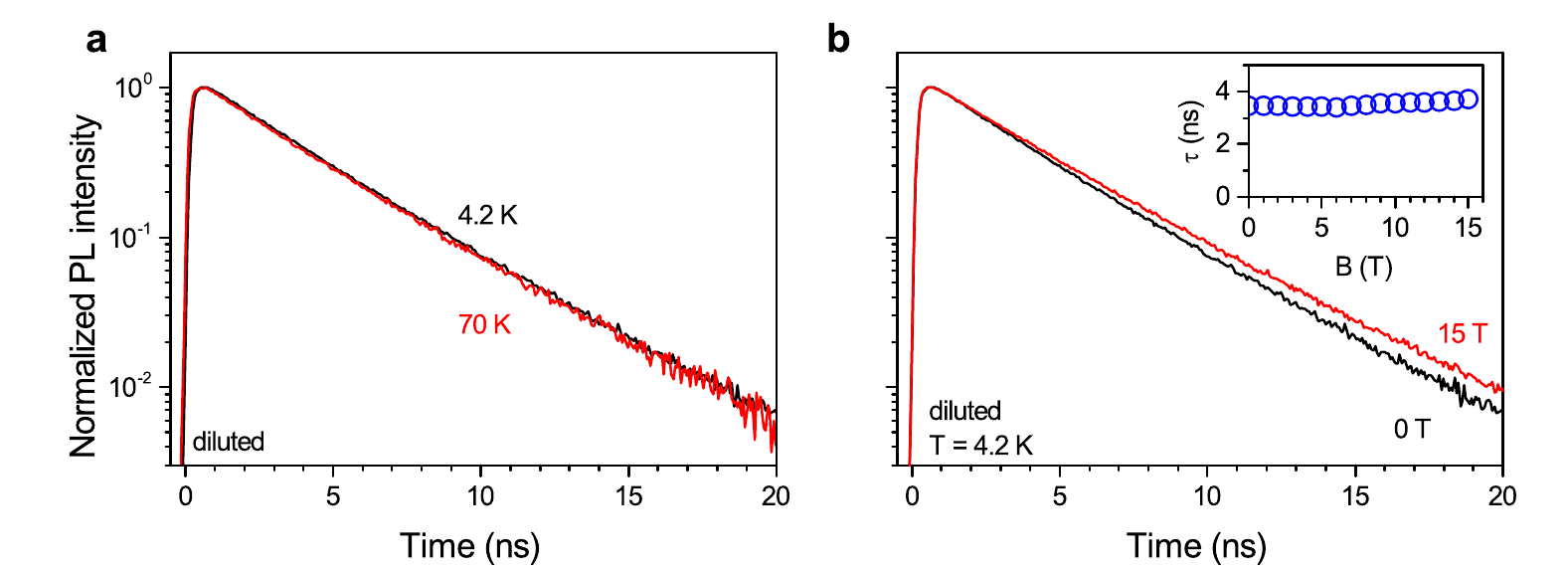}
    \caption{(a) PL decays of a diluted dropcasted CdSe/CdS NPL sample at $T=4.2$~K (black) and 70~K (red). (b) PL decays at $B=0$ (black) and 15~T (red), $T=4.2$~K. Inset: PL decay time dependence on magnetic field.
}
\label{fig:SI_TR(T,B_diluted)}
\end{figure*}

\subsection{S3. Statistics for nanoplatelets with various dilutions}

We use magneto-optical technique and DCP analysis to study statistics of the NPL orientation in the dropcased samples prepared from solutions with different dilutions. Four solutions with the following dilutions were investigated: sample 1 (dense), sample 2 (dilution $1:10$), sample 3 ($1:200$) and sample 4 ($1:800$).  It is typical for dropcasted samples that they are not homogeneous and NPL concentration and, respectively, NPL orientations may vary from point to point. Therefore, statistical approach is required here. For each sample we have measured 20 points moving sample by a piezostage. The inspected area was about 3~mm$^2$ and the laser focus spot for each measurement was about 100~$\mu$m in diameter. DCP was measured at a magnetic field of 15~T at $T=4.2$~K. In order to speed up data collection we measured time-integrated DCP, $P_c^{\rm int}$, with using CCD for PL detection. As we show in Section S4, in the studied CdSe/CdS NPLs with the fast spin relaxation of trions $P_c^{\rm int}$ does not differ much from $P_c^{\rm eq}$.

\begin{figure*}[h!]
    \includegraphics[width=16cm]{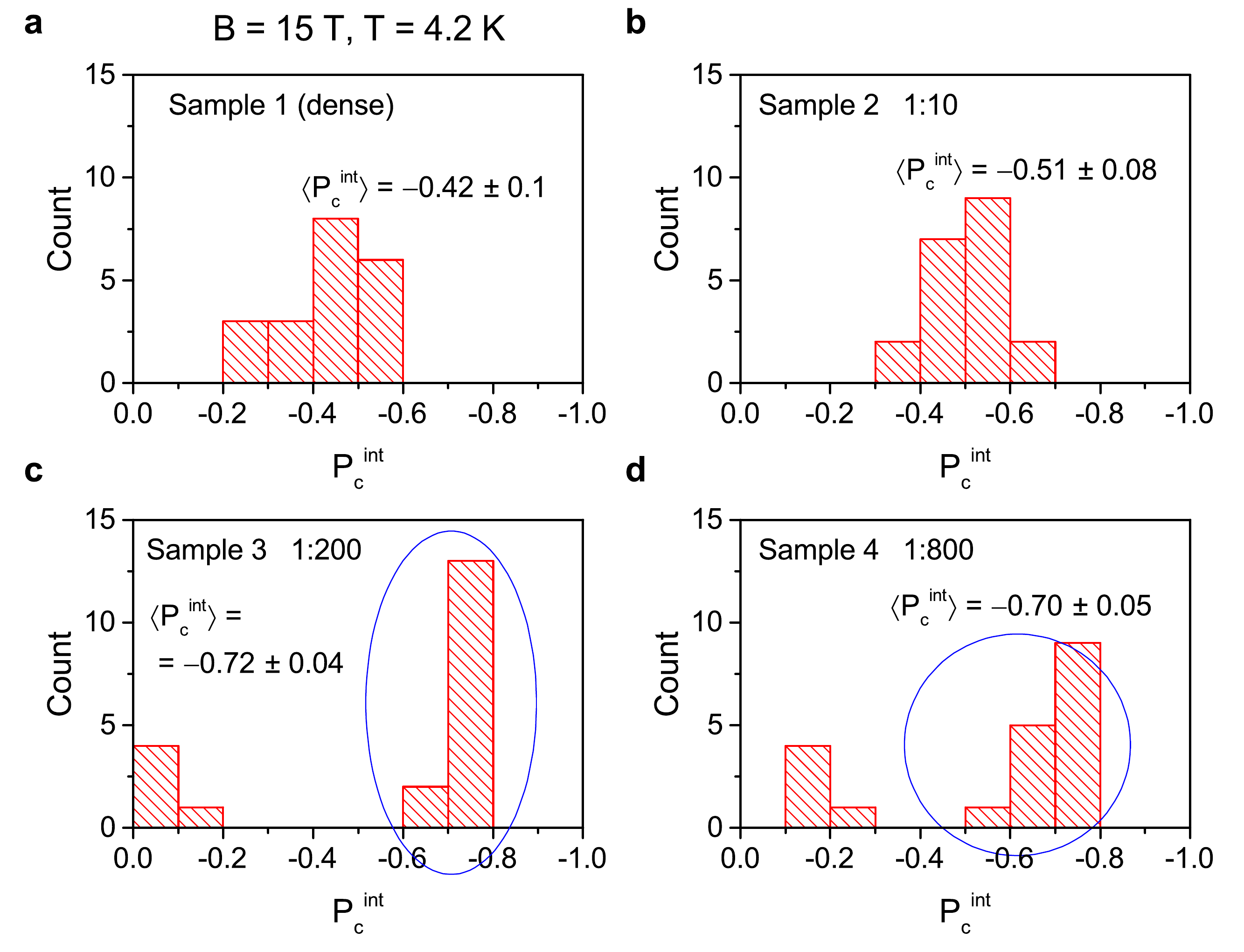}
    \caption{$P_c^{\rm int}$ occurrence in the dense and diluted ensembles of NPLs determined from 20 measurements at different sample positions (20 counts for each sample).}
    \label{fig:SI_DCP_different_spots_statistics}
\end{figure*}

\begin{figure*}[h!]
    \includegraphics{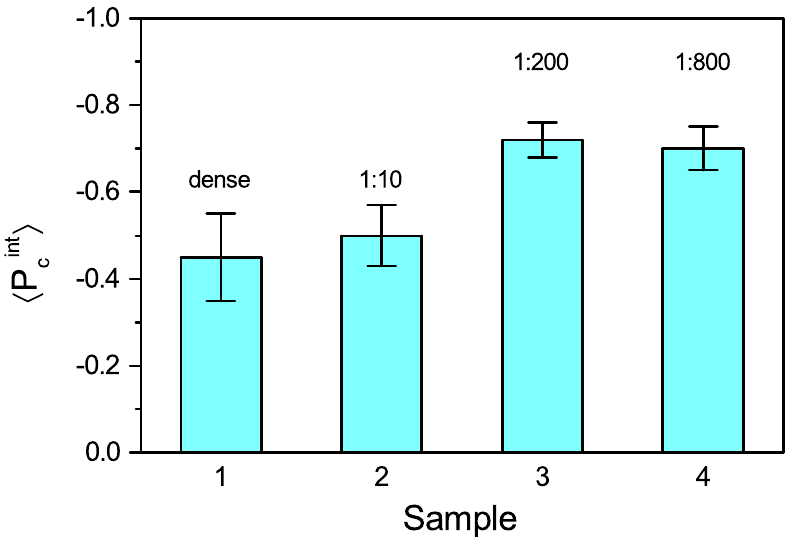}
    \caption{Averaged DCP in NPL ensembles with various dilutions. }
    \label{fig:SI_DCP_histogram}
\end{figure*}

Experimental results shown in Figure~\ref{fig:SI_DCP_different_spots_statistics} demonstrate that in the dense sample the averaged time-integrated DCP, $\langle P_c^{\rm int} \rangle = \sum\limits_{m=1}^n (P_c^{\rm int})_m / n$, where $n=20$ in the number of measured points, $(P_c^{\rm int})_m$ is time-integrated DCP in the $m$th point, equals to $-0.42$. It increases only slightly to $-0.51$ for $1:10$ dilution, but then more drastically to $-0.72$ for dilutions of $1:200$ and $1:800$. These results are summarized as a histogram in Figure~\ref{fig:SI_DCP_histogram}. It is obvious, that the NPL dilution affects on the ratio between horizontally and vertically oriented NPLs in the dropcasted samples: in the diluted samples horizontal orientation dominates.

\newpage

\subsection{S4.  Comparison of equilibrium and time-integrated DCP}

It is instructive to compare the equilibrium DCP values $P_c^{\rm eq} \equiv P_c(t \to \infty)$ with the time-integrated values $P_c^{\rm int}$, as the latter corresponds to the values recorded in commonly performed measurements under continuous-wave excitation and/or detection without time resolution. Modeling consideration for that is developed in Refs.~\cite{Liu2013,Liu2014,Siebers2015}. Let us remind here some basic points. We are considering trions here, while the approach is general and can be equally used for excitons.

The degree of circular polarization (DCP) of photoluminescence is defined by
\begin{equation}
P_c(t)= \frac{I^+(t) - I^-(t)}{I^+(t) + I^-(t)}  .
\label{eq:DCP_1}
\end{equation}
Here, $I^+(t)$ and $I^-(t)$ are $\sigma^+$ and $\sigma^-$ circularly polarized PL intensities,
respectively, measured at a time delay $t$ after pulsed excitation. A saturation of $P_c(t)$ at
times much longer than the trion spin-relaxation time gives the equilibrium circular polarization
degree $P_c^{\rm eq}(B)$.

The time-integrated DCP can be evaluated by integrating the corresponding PL intensities over time:
\begin{equation}
P_c^{\rm int}= \frac{\int dt I^+(t) - \int dt I^-(t)}{\int dt I^+(t) + \int dt I^-(t)}  .
\label{eq:DCP_2}
\end{equation}
In the case of continuous-wave excitation, the measured polarization degree corresponds to $P_c^{\rm int}$.

The magnetic-field-induced DCP is caused by trion thermalization on the Zeeman spin levels. While
the equilibrium polarization degree $P_c^{\rm eq}(B)$ is controlled solely by the thermal
equilibrium population of the Zeeman spin levels, the time-integrated polarization $P_c^{\rm int}$
depends also on the ratio of  the trion spin relaxation time $\tau_s$ to the trion lifetime $\tau$:
\begin{equation}
P_c^{\rm int}(B)= \frac{\tau}{\tau + \tau_s}P_c^{\rm eq}(B)  .
\label{eq:DCP_3}
\end{equation}
If $\tau_s \ll \tau$, the experimentally measured $P_c^{\rm int}(B)$  coincides with $P_c^{\rm
eq}(B)$, while otherwise their difference is controlled by the dynamical factor $d = \tau/(\tau +
\tau_s)$. It is important to note, that the magnetic field dependence of $P_c^{\rm int}(B)$ can be
rather complicated as $\tau_s$ and  $\tau$ can also be functions of magnetic field. Moreover, for
the negatively charged trions, where Zeeman splitting is controlled by the heavy-hole $g$-factor,
$\tau_s(B)$ dependence becomes sensitive to the NPL orientation in respect to magnetic field
direction, i.e., is $\tau_s(B, \theta)$. Rigorous account of this effect for the ensemble
of randomly oriented nanostructures is suggested in Eq.~(9) of Ref.~\onlinecite{Liu2013}.

In the present studies of the ensemble of the thick-shell NPLs we assume that their
anisotropic axis may be oriented only horizontally ($\theta=0$)  or vertically ($\theta=\pi/2$)
with respect to the magnetic field applied in the Faraday geometry. As we discuss in detail in the
next section S5, mostly horizontally oriented NPLs contribute to the DCP. Therefore, the
experimentally measured spin relaxation time from Figure~\ref{fig:TR}b directly corresponds to $\tau_s(B)$ for the magnetic field orientation along NPL axis. Note, that in this respect the case of thick-shell NPLs is simple for interpretation and model treatment, e.g. compared to ensemble of randomly oriented QDs, where the angle dependence of $\tau_s(B, \theta)$ need to be taken into account.\cite{Liu2013}

\begin{figure}[h!]
    \includegraphics[width=16cm]{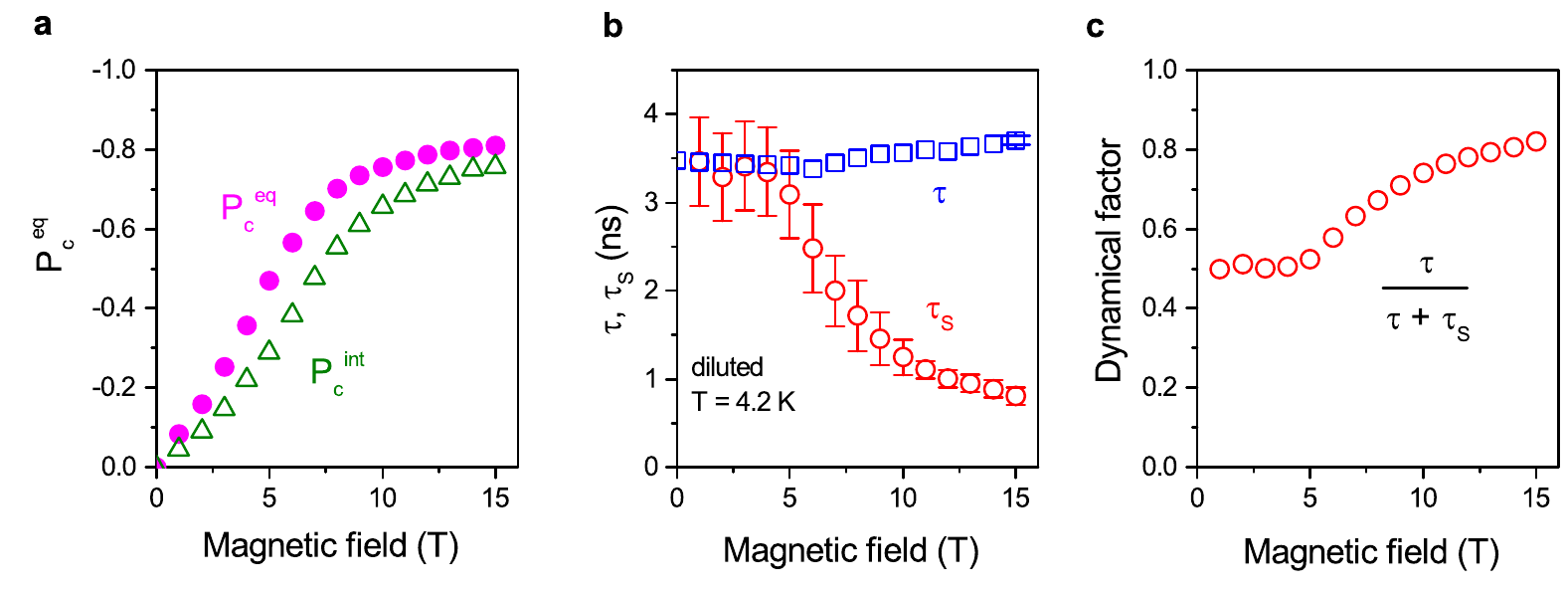}
        \caption{DCP results for the diluted NPL ensemble. $T=4.2$~K. (a) Comparison of the equilibrium and time-integrated DCP. (b) Magnetic field dependences of the trion lifetime and trion spin relaxation time. (c) Magnetic field dependence of the dynamical factor $d = \tau/(\tau + \tau_s)$.
        }
\label{fig:Dyn_factor}
\end{figure}

Experimental results for $P_c^{\rm int}$ and $P_c^{\rm eq}$ measured for the diluted NPL ensemble are shown in Figure~\ref{fig:Dyn_factor}a. Indeed, difference between them is not very large. Magnetic field dependences of the trion lifetime and trion spin relaxation time are shown in Figure~\ref{fig:Dyn_factor}b and the field dependence for the dynamical factor $d = \tau/(\tau + \tau_s)$ in Figure~\ref{fig:Dyn_factor}c.

\newpage

\subsection{S5.  Modeling of DCP in thick-shell nanoplatelets}

The DCP is determined by the difference between intensities of $\sigma^+$ and $\sigma^-$ polarized
light ($I^{+}(\theta)$ and $I^{-}(\theta)$) in ensemble of NPLs with anisotropy axis oriented at
the angle $\theta$ to the direction of magnetic field. Intensities $I^{\pm}(\theta)$ depend on two
quantities: (i) within the  electric-dipole approximation the relative probabilities of detecting
$\sigma^\pm$ light from negative trion in NPL oriented at an angle $\theta$ to the observation
direction are $\chi_{3/2}(\sigma^\pm)\propto(1\pm\cos\theta)^2$ and
$\chi_{-3/2}(\sigma^\pm)\propto(1\mp\cos\theta)^2$; (ii) equilibrium population of trion spin
levels $\pm 3/2$ separated by the Zeeman splitting $\Delta E_{\rm hh}=E_{\rm hh}^{+3/2}-E_{\rm
hh}^{-3/2}=- 3g_{h}\mu_{\rm B}B\cos\theta = - 3g_{h}\mu_{\rm B}Bx  $, where $x=\cos\theta$.
However, the recombination rates of the negative trions are not influenced by magnetic
field and thus do not depend on the NPL orientation. As result, intensities of $\sigma^+$ and
$\sigma^-$ polarized light are:
\begin{equation}
I_{\sigma^{\pm}}(x) \propto \frac{(1\mp x)^2 \exp(\Delta E_{\rm hh}/2kT)+(1\pm x)^2 \exp(-\Delta E_{\rm hh}/2kT)}
{[\exp(\Delta E_{\rm hh}/2kT)+\exp(-\Delta E_{\rm hh}/2kT)]}.
\end{equation}
In general case the spatial orientation of NPLs in ensemble can be described by the
weighting function $f_{\rm or}(x)$. The equilibrium value of DCP, $P_c^{\rm eq}(B)$, in this case
is described by the following equation:\cite{Siebers2015}
\begin{equation}\label{Pcgen}
P_c^{\rm eq}(B)=\frac{\int_{0}^{1}f_{\rm or}(x)[I^{+}(x)-I^{-}(x)]dx}{\int_{0}^{1}f_{\rm or}(x)[I^{+}(x)+I^{-}(x)]dx}= -
\frac{\int_{0}^{1}f_{\rm or}(x)2x\tanh(\Delta E_{\rm hh}/2kT)dx}{\int_{0}^{1}f_{\rm or}(x)(1+x^2)dx},
\end{equation}
 Distribution function $f_{\rm or}(x)=1$ corresponds to the ensemble of
randomly oriented NPLs and gives maximum achievable $|P_c^{\rm sat}|=|P_c^{\rm eq}(B\rightarrow
\infty)|=0.75$, which is usually observed for spherical CdSe QDs. Here, for NPLs we consider  the
case of the bimodal distribution, when NPLs are oriented only in horizontal and vertical
directions: $f_{\rm or}(x)=n_{\rm hor} \delta (1-x) + n_{\rm ver} \delta (x)$. Here $n_{\rm hor}$ and $n_{\rm vert}$ are the fractions of NPLs lying horizontally or vertically, respectively. By definition $n_{\rm hor}+n_{\rm vert}=1$. With such distribution Eq.~(\ref{Pcgen}) reduces to
\begin{equation}
P_c^{\rm eq}(B)= \frac{2n_{\rm hor} }{2n_{\rm hor}+n_{\rm ver}}   \tanh \frac{3g_{h}\mu_{\rm B} B}{2kT} .
\label{Psat}
\end{equation}
One can see, that maximum achievable $P_c^{\rm eq}$ in this case equals $|P_c^{\rm sat}|=2n_{\rm
hor}/(2n_{\rm hor}+n_{\rm vert})=2n_{\rm hor}/(n_{\rm hor}+1)$.

\begin{figure}[h!]
    \includegraphics[width=10cm]{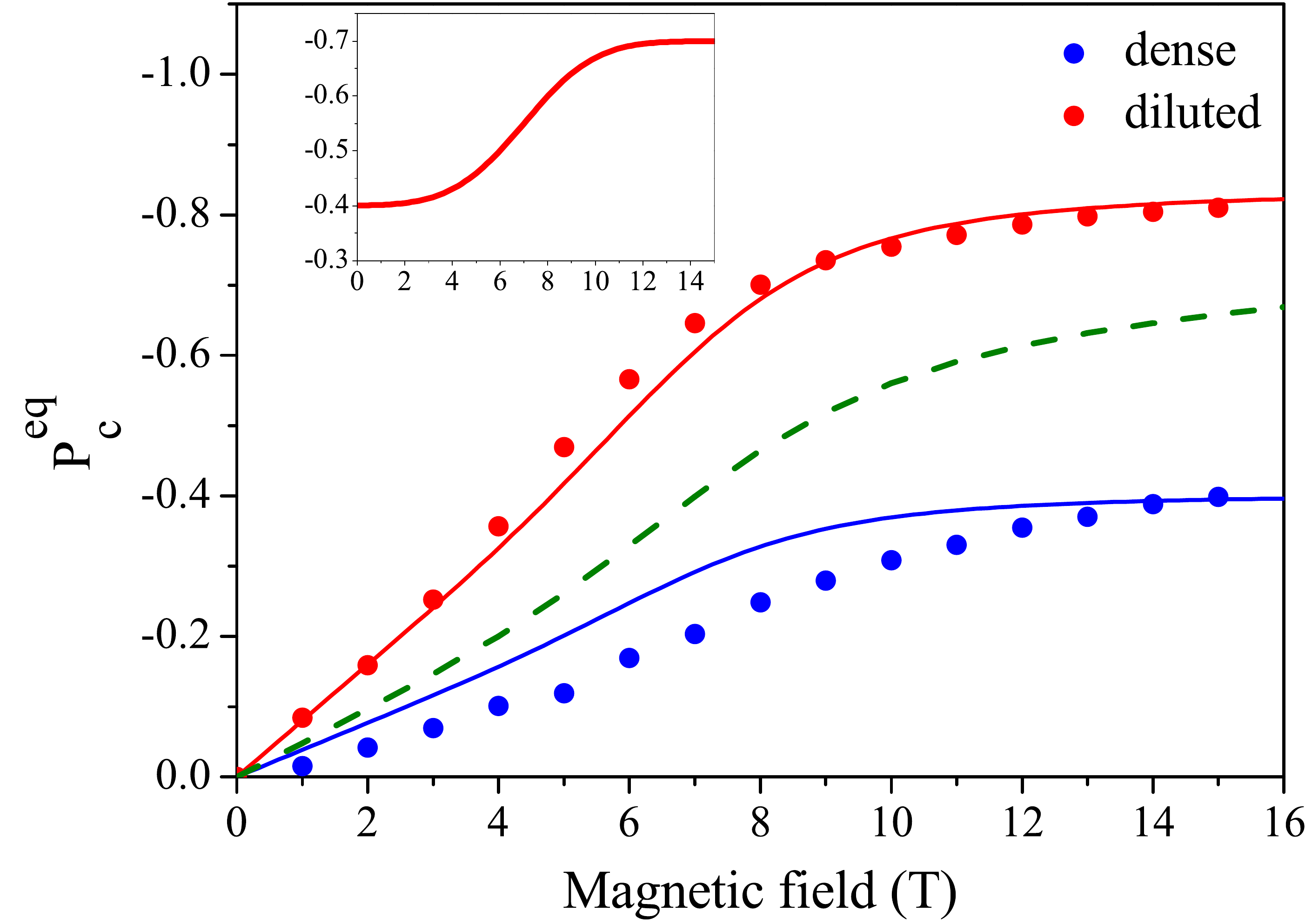}
        \caption{Modeling of the $P_c^{\rm eq}$ in diluted (red line) and dense (blue line) ensembles of NPLs with bi-modal orientations and in the randomly orientated NPL ensemble (dashed green line). Experimental data are shown by symbols. Magnetic field dependence of the hole $g$-factor is shown in the insert.
        }
\label{Fig2}
\end{figure}

Figure~\ref{Fig2}  shows the modeling   of $P_c^{\rm eq}(B)$ with $n_{\rm hor} =0.70$ and $0.25$  for diluted and  dense ensembles, correspondingly, and magnetic-field-dependent hole $g$-factor. Green curve shows  $P_c^{\rm eq}(B)$ for the randomly oriented ensemble with the same magnetic-field-dependent hole $g$-factor.

\begin{figure}[h!]
    \includegraphics[width=10cm]{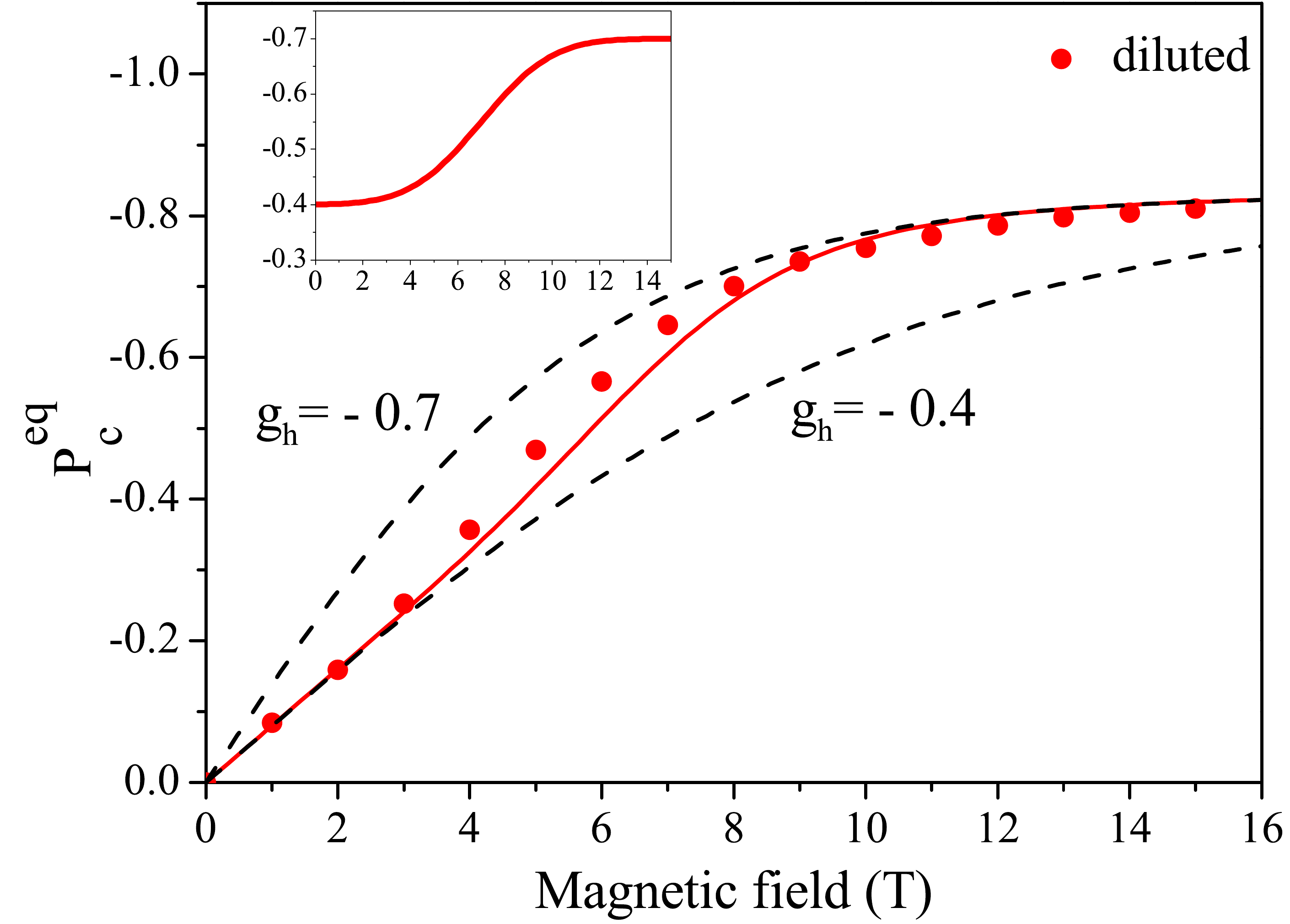}
        \caption{Modeling of the $P_c^{\rm eq}(B)$ experimental data for in the diluted NPL ensemble with the magnetic-field-dependent hole $g$-factor (red line) shown in the insert and for two constant hole $g$-factors $g_h=-0.4$ and $-0.7$ (dashed lines).}
\label{Fig3}
\end{figure}

\begin{figure}[h!]
    \includegraphics[width=10cm]{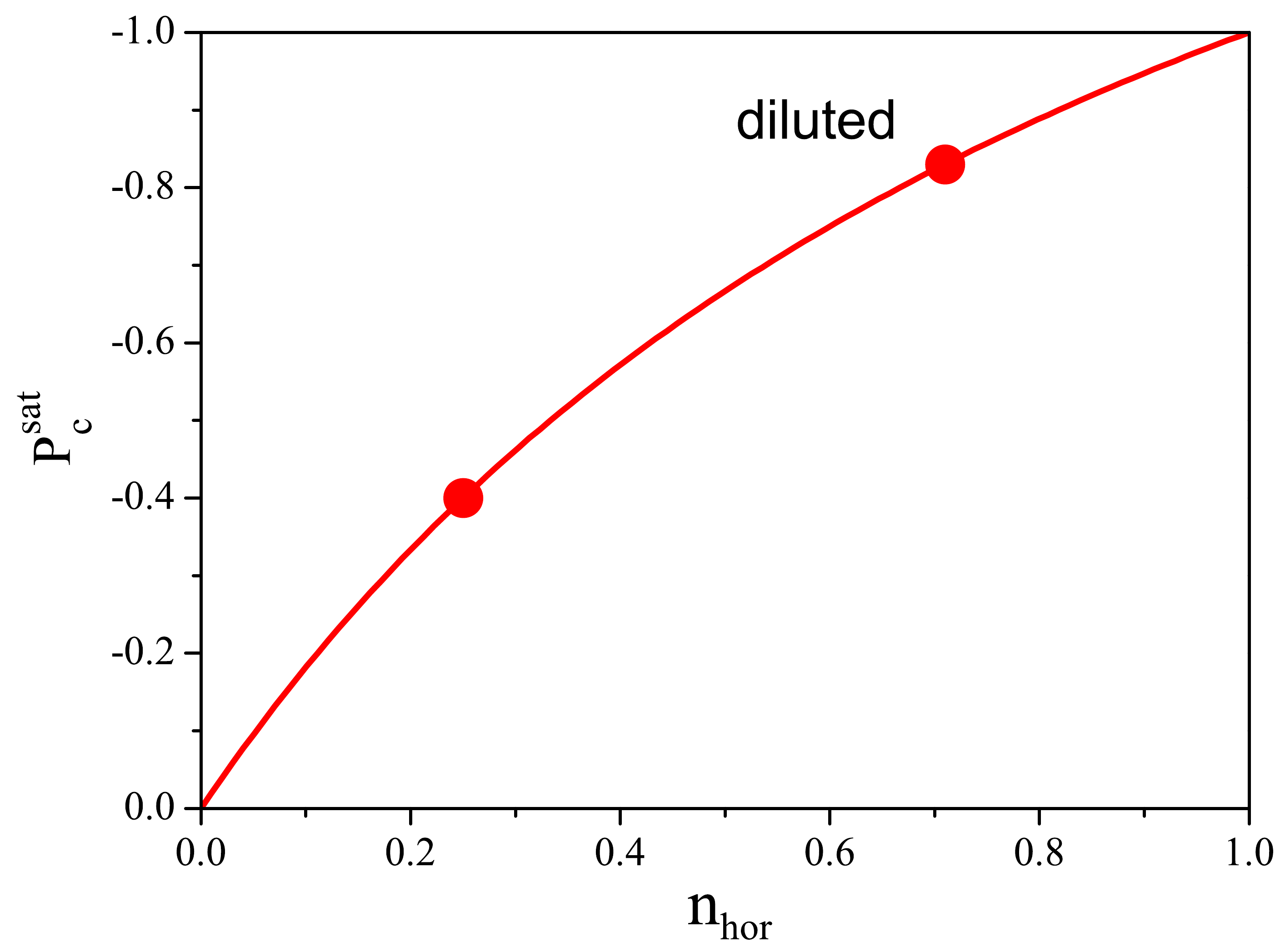}
        \caption{Saturation DCP in strong magnetic fields vs. fraction of the horizontally oriented NPLs. The dependence is calculated with $P_c^{\rm sat}= - 2n_{\rm hor}/(n_{\rm hor}+1)$.  Marked points on the dependence correspond to the dense and diluted samples from Figure~\ref{fig:TR}.   }
\label{Fig4}
\end{figure}

In Figure~\ref{Fig3} we compare the modeling of the $P_c^{\rm eq}(B)$ experimental data in the diluted NPL ensemble with the magnetic-field-dependent hole $g$-factor (red line) and for two constant hole $g$-factors $g_h=-0.4$ and $-0.7$. In Figure~\ref{Fig4} the calculated dependence of the $P_c^{\rm sat}$ on the fraction of horizontal NPLs $n_{\rm hor}$ is presented. It allows us to estimate the preferable orientation of NPLs on the substrate by optical means. The only needed experimentally measured value for this evaluation is $P_c^{\rm sat}$, while in colloidal structures it may require using of strong magnetic fields exceeding 10~T.

\newpage

\subsection{S6. Calculation of hole $g$-factor in nanostructures}

We calculate the hole $g$-factor for semiconductor nanostructures with large spin-orbit
splitting, which allows us to neglect the spin-split valence subband. The hole is at the top of the
four times degenerate $\Gamma_8$ valence subband and its Hamiltonian can be written  as
\begin{equation}\label{Hamilt}
\widehat{H}=\widehat{H}_L+\widehat{H}_Z+\widehat{H}_B+V_{\rm ext}({\bf r}).
\end{equation}
Here
\begin{equation}\label{lutt}
\widehat{H}_L=\frac{\hbar^2}{2 m_0}\left[\left(\gamma_1+\frac{5}{2}\gamma\right)k^2-2\gamma({\bf kJ})^2\right],
\end{equation}
is the Luttinger Hamiltonian in zero magnetic field in spherical approximation
\cite{Luttinger,Gelmont71}, where the valence band warping is neglected.   For the $g$-factor
estimations the spherical approximation is sufficient.  ${\bf J}$ is the hole internal angular
momentum operator for $J=3/2$, $m_0$ is the free electron mass,  $\gamma_1$ and
$\gamma=(2\gamma_2+3\gamma_3)/5$ are Luttinger parameters related to the light-hole and heavy-hole
effective masses as $m_{\rm l,h} = m_0/(\gamma_1 \pm 2 \gamma)$ (with $+$ for light hole and $-$ for heavy hole), $V_{\rm ext}({\bf
r})$ is the quantum structure potential.

External magnetic field  $\bf B$  contributes to the hole Hamiltonian \eqref{Hamilt},
firstly, inducing the Zeeman splitting \cite{Roth,Luttinger}
\begin{equation}
\label{Zbulk}
\widehat{H}_Z=-2\mu_{\text{B}}\varkappa \left(\bf J\bm B\right),
\end{equation}
where $\mu_{\text{B}}$ is the Bohr magneton, $\varkappa$ is the Luttinger magnetic constant.  The
value of $\varkappa$ can be estimated by perturbation theory via other Luttinger parameters as
\cite{Roth}:
\begin{equation}
\label{kappa2}
\varkappa\approx-2/3+5\gamma/3-\gamma_1/3 .
\end{equation}
Cubic symmetry contribution to the Zeeman effect is $\propto q(B_x J_x^3+B_y J_y^3 + B_z J_z^3)$,
 where $J_x$, $J_y$ and $J_z$ are the matrices of the projections of the operator ${\bf
J}$ on the crystallographic axes in the basis of the Bloch functions $u_\mu$ ($\mu=\pm 3/2, \pm
1/2$) for the top of the valence band. This contribution is described by the anisotropic parameter
$q$ which in most structures is small as compared with the contribution of the isotropic parameter
$\varkappa$ in equation \eqref{Zbulk} \cite{X_Marie}, therefore, in the following estimations it is
neglected.

Secondly, the magnetic field induced Hamiltonian $\hat H_B$ describes the orbital contribution and
can be obtained by the replacement in the Luttinger Hamiltonian of the operator $\bf k$ with
$\mathbf k -\frac{e}{c}\bf A$, where $\bf A$ is the vector potential of the magnetic field. Here we
use $\gamma_1,\gamma>0$, $e=|e|$ and chose the $z$ axis along the direction of the magnetic field.
The vector potential is chosen in Landau gauge, $\bf A=(0,Bx,0)$. We take as the energy and length
units the effective Rydberg, $Ry=m_{h} e^4/2 \ \hbar^2$, and Bohr radius,
$a_{\text{B}}=\hbar^2\/m_{h} e^2$. In these units the dimensionless  hole Hamiltonian can be
written as
\begin{equation}\label{luttB}
\widehat{H}=\frac{2\gamma_1+5\gamma}{2(\gamma_1-2\gamma)}k^2-\frac{2\gamma}{\gamma_1-2\gamma}({\bf
kJ})^2+\tilde{V}_{\text{ext}}({\bf r})+\widehat{\tilde{H}}_Z+\widehat{\tilde{H}}_B
.\end{equation} Here $\widehat{\tilde{H}}_Z=-\frac{2\varkappa}{\gamma_1-2\gamma}\tilde{B}J_z$
is the dimensionless Zeeman Hamiltonian
\eqref{Zbulk}, $\widehat{\tilde H}_B$ is the dimensionless Hamiltonian describing the orbital
contribution of the magnetic field:
\begin{equation}\label{HB}
\widehat{\tilde{H}}_B=\frac{2\gamma_1+5\gamma}{2(\gamma_1-2\gamma)}\left(-2\tilde{B}xk_y+\tilde{B}^2x^2\right)+
\end{equation}
$$+\frac{2\gamma}{\gamma_1-2\gamma}\left[J_y^2\left(2\tilde{B}xk_y-\tilde{B}^2x^2\right)
+\left(J_yJ_z+J_zJ_y\right)\tilde{B}xk_z+\left(J_yJ_x+J_xJ_y\right) \frac{\tilde{B}}{2}(k_x x+x k_x)\right],$$
where $\tilde{B}=\frac{e a_B^2}{c\hbar}  B$ is the dimensionless magnetic field.
 In $\widehat{\tilde{H}}_{\text{B}}$ the quadratic contributions $\propto B^2$ are taken into account,
although they do not contribute to the hole effective $g$-factor, which is defined in the low field
limit and is determined by linear in magnetic field terms.

In the limit of weak magnetic fields the spin splitting of the degenerate in
zero magnetic field energy levels $\Delta E$ is linear in the field strength. That allows us to
introduce the effective hole $g$-factor as
\begin{equation}
\label{g:def}
g_h = \frac{E_{-|M|}-E_{|M|}}{2|M|\mu_{\text{B}} B}\, ,
\end{equation}
where $E_M$ is the energy of the hole with the  projection $M$ of the hole total momentum ${\bf J}
+ {\bf L}$ (where ${\bf L}=\hbar [{\bf r} \times {\bf k}]$ is the external orbital momentum) on the
magnetic field direction. For the lowest heavy-hole states $M=\pm 3/2$ and for the lowest light-holes
states $M=\pm 1/2$. The $g$-factor sign is defined  so that for $g_h>0$ the lowest hole
state is the state with positive $M>0$ according to the definition in Refs.~\onlinecite{Efros1996,Efros2003}

 The value of the hole effective $g$-factor depends on the hole wave function,
and, therefore, on the external confining potential $V_{\rm ext}$ acting on the hole. It is
possible to make analytical estimations for the hole $g$-factor for several types of $V_{\rm ext}$.
We consider the spherically symmetric  $V_{\rm ext}$ for the quantum dots with abrupt (box-like)
and smooth  spatial confinement, as well as the 2D thin quantum wells with the abrupt and smooth
confinement along the direction of the anisotropic axis parallel to the magnetic
field direction. The estimations of the hole $g$-factors in CdSe-based structure with four different $V_{\rm ext}$ are shown in Table~\ref{Table_g1} in the main text and for other II-VI semiconductor nanostructures in Table~\ref{Table_g2}.  As it was noted, the  $g$-factor of the hole moving in the external potential depends on the hole wave functions, and, therefore, the specific values of the Luttinger parameters are important.  We examined several different parameterizations for each semiconductor taken from the literature
\cite{Adachi, Karazhanov, Horodyska} and considered materials with both cubic and wurtzite crystal
structures. Resulting $g$-values shown in Tables~\ref{Table_g1} and \ref{Table_g2} are obtained as follows:

1. Column $g_{\rm bulk}$ shows  the hole $g$-factor at the top of the valence band in the bulk
semiconductor in the absence of the external potential:  $g_h=g_{\rm bulk}=2\varkappa$. It comes
solely from the Zeeman Hamiltonian $\hat H_Z$.

2. In spherical quantum dots the lowest hole ground state is four-fold degenerate.  In magnetic field this state splits into four equidistant levels as $M g_h\mu_{\rm B}B$ (where $M=\pm 3/2,\pm 1/2$) with effective $g$-factor  $g_h=g_{\rm sph}$:   \cite{Gelmont73}
\begin{equation}\label{Gelmont}
g_{\rm sph}   = 2\varkappa+\frac{8}{5}\gamma I_1^{\text{g}}+\frac{4}{5}\left[\gamma_1-2\left(\gamma+\varkappa\right)\right]I_2^{\text{g}} \, ,
\end{equation}
where \begin{equation}\label{Gelmont1} I_1^{\text{g}}=\int\limits_{0}^{\infty}r^3 R_2(r) \frac{d
R_0(r)}{ dr} dr,~ I_2^{\text{g}}=\int\limits_{0}^{\infty}r^2 R_2^2(r) dr \, .
\end{equation}
Here $R_0(r)$ and $R_2(r)$ are the hole radial functions which   are normalized as $\int (
R_0^2 + R_2^2 ) r^2 dr  =1$ and satisfy the system of radial equations from Ref.\cite{Gelmont71}.
The integrals $I_1^{\text{g}}$ and $I_2^{\text{g}}$  contain the same power of radial functions and
hole coordinate as the normalization integrals. Therefore, they are independent of the quantum dot
size and  depend only on the type of the confining potential and on the light-hole to heavy-hole
effective mass ratio, $\beta=(\gamma_1-2\gamma)/(\gamma_1+2\gamma)$. The resulting hole $g$-values
depend on the effective mass parameters in the specific semiconductor,  type of the dot potential, but are independent of the QD size.

Columns $g_{\text{sph}}^{(\text{par})}$ and  $g_{\text{sph}}^{(\text{box})}$ show the hole
effective $g$-factors calculated with the hole radial wave functions in  the smooth and abrupt
(box-like) potentials for deeply confined states, respectively.  For the smooth potential the parabolic-like $V_{\rm ext} \propto  r^2$ was chosen. It was shown that  the hole effective $g$-factor is almost the same in parabolic-like and  Gaussian-like potentials and is independent of the potential parameters. \cite{Semina2016} For the most of valence band parametrizations, the $g$-factors fall in the range from $-0.8$ to $-1.2$ independent of the potential type and the semiconductor crystal
structure.

3. In thin quantum well-like structures like NPLs, where the width along the anisotropic axis
directed along the magnetic field is much smaller than other dimensions, the effective hole
$g$-factor can be well approximated as one in the thin quantum well. In this case the heavy-holes
and light-holes are split and the ground state is the heavy hole. The $g$-factor is determined by
the magnetic field induced heavy-hole and light-hole subband mixing. The mixing is mediated by the
nonzero wave vector $k_z$ defined by the quantization along the anisotropic axis and by the nonzero
in-plane component of  $\mathbf \hbar k -\frac{\hbar e}{c}\bf A$. It can be calculated in the second order of the perturbation theory so that the resulting $g_h=g_{\rm QW}$ is given by \cite{Wimbauer}:
\begin{equation}\label{g2d}
g_{\text{QW}} =2\varkappa- 4\frac{\hbar^2}{m_0}\sum\limits_{n=1}^\infty \frac{|\langle lh_{2n}|\gamma \hat{k}_z|hh_1\rangle|^2}{E_{lh_{2n}}-E_{hh_1}},
\end{equation}
where $|hh_1\rangle$ is the wave function of the heavy-hole  ground state of the quantization along
$z$-axis with zero wave vector in structure plane, $|lh_{2n}\rangle$ is the even excited states of
the light hole, $E_{hh_1}$ and  $E_{lh_{2n}}$  are the corresponding energies. Note, that
the numerator and denominator of the perturbation theory series in Eq.~\eqref{g2d} scale with the QW
width $L$ as $L^{-2}$ resulting in hole $g$-factor independence of the QW width. At the same time, the
$g$-factor remains very sensitive to the type of confining potential.  The hole $g$-factors in
two-dimensional structure with smooth (parabolic) and box-like potentials are shown in columns
$g_{\text{QW}}^{(\text{par})}$ and $g_{\text{QW}}^{(\text{box})}$, correspondingly. The $g$-factors
in thin structures are strongly dependent on both the confining potential and the parametrization
for the cubic semiconductors. In most cases the absolute values of the $g$-factors are smaller than
that for the spherical quantum dots. In contrast, the $g$-factors for the wurtzite materials  are
almost in the same region as for the spherical quantum dots.


In the presence of the other charge carrier, e.g. for the hole in the exciton, the calculation of the hole $g$-factor becomes more
complicated and the simple estimations in most cases are not possible. It can be demonstrated, that
in the limit of the quantum well-like structure the $g$-factors are the same as for two-dimensional
isolated hole.  The condition is the exciton binding energy being much smaller than the distances
between several low lying hole energy levels of size quantization along growth axis. If it is not the
case, the hole $g$-factor calculation turns to the complicated problem, which is outside the scope
of this paper.

\begin{figure}
    \includegraphics[width=16cm]{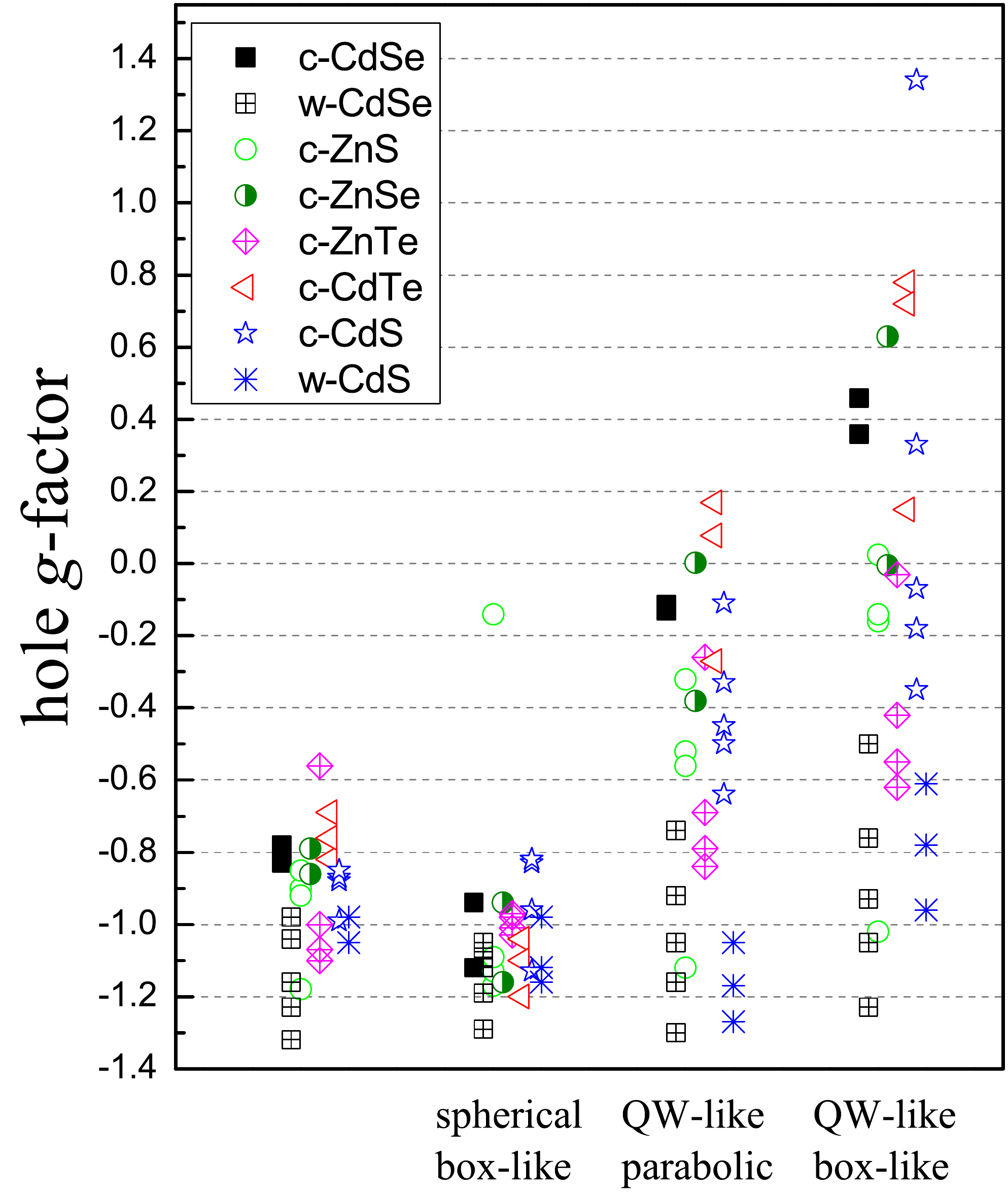}
    \caption{Calculated effective hole $g$-factors for semiconductor nanostructures.}
\label{gfact}
\end{figure}

The calculated hole effective $g$-factors from Table~\ref{Table_g1} and Table~\ref{Table_g2}  are shown in
Figure~\ref{gfact}. To summarize, the hole $g$-factors in bulk materials calculated for two types of confining potentials  demonstrate the strong dependence on the potential type (smooth or the box-like) and on the parametrization (particular values of the Luttinger parameters)  for the quantum well-like
structures. Therefore, one has to be careful in choosing the model potential and the material
parameters for the quantum well-like structures. In opposite, for the spherically symmetric structures these dependencies are much weaker.

\begin{table}
    \begin{center}
         \caption{Hole $g$-factors calculated for cubic and wurtzite semiconductor nanostructures.}
        \label{Table_g2}
        \begin{tabular}{| l | l | l | l |  l | l | l |  l | l |  l |l |}
            \hline
            Material & $\gamma_1$ & $\gamma$ & $\varkappa$&$g_{\rm bulk}$ &$\beta$&$g_{\text{sph}}^{(\text{par})}$&$g_{\text{sph}}^{(\text{box})}$
            &$g_{\text{QW}}^{(\text{par})}$&$g_{\text{QW}}^{(\text{box})}$&Refs.\\ \hline
            c-ZnS& 1.77 & 0.492 & -0.437&0.87 &0.28&-1.18&-1.17&-1.12&-1.02&\onlinecite{Dietl} \\ \hline
            c-ZnS& 2.54 & 0.954 & 0.077&0.15&0.14&-0.9&-1.13&-0.52&-0.16&\onlinecite{Lawaetz} \\ \hline
            c-ZnS& 3.158 & 1.0556 & 0.04&0.08 &0.2&-0.92&-1.09&-0.56&-0.25&\onlinecite{Karazhanov} \\ \hline
            c-ZnS& 2.98 & 1.156 & 0.26&0.53 &0.126&-0.85&-1.14&-0.32&-0.14&\onlinecite{Karazhanov} \\ \hline
&  &  & &  &  &   &  &  &   &   \\ \hline
            c-ZnSe& 3.94 & 1.312 & 0.207&0.413 &0.2&-0.86&-0.94&-0.38&-0.004&\onlinecite{Adachi} \\ \hline
            c-ZnSe&3.77 & 1.5 & 0.57&1.14 &0.11&-0.79&-1.16&0.003&0.63&\onlinecite{Lawaetz} \\ \hline
&  &  & &  &  &   &  &  &   &   \\ \hline
            c-ZnTe& 4 & 1.112 & -0.147&-0.29 &0.28&-0.56&-1.03&-0.26&-0.03&\onlinecite{Stradling} \\ \hline
            c-ZnTe& 3.8 & 1.136 & -0.04&-0.08 &0.25&-1&-0.98&-0.69&-0.42&\onlinecite{Said} \\ \hline
            c-ZnTe& 3.9 & 1.112 & -0.11&-0.22 &0.27&-1.07&-0.97&-0.79&-0.55&\onlinecite{Oka}\\ \hline
            c-ZnTe& 3.8 & 1.068 & -0.15&-0.306 &0.28&-1.1&-1.01&-0.84&-0.62&\onlinecite{Wagner} \\ \hline
&  &  & &  &  &   &  &  &   &   \\ \hline
            c-CdTe& 4.14 & 1.4 & 0.3&0.6 &0.19&-0.82&-1.04&-0.27&0.15&\onlinecite{Friedrich} \\ \hline
            c-CdTe& 5.3 & 1.88 & 0.7&1.4 &0.17&-0.69&-1.1&0.17&0.78&\onlinecite{Said}\\ \hline
            c-CdTe& 4.11 & 1.6 & 0.63&1.26 &0.12&-0.76&-1.2&0.078&0.72&\onlinecite{Neumann} \\ \hline
&  &  & &  &  &   &  &  &   &   \\ \hline
            c-CdS& 4.11 & 1.226 & 0.007&0.014 &0.25&-0.99&-0.96&-0.64&-0.35&\onlinecite{Adachi} \\ \hline
            c-CdS& 2.721 & 1.027 & 0.138&0.27 &0.14&-0.88&-1.13&-0.45&-0.07&\onlinecite{Karazhanov}\\ \hline
            c-CdS& 2.647 & 1.098 & 0.28&0.56 &0.093&-0.87&-0.82&-0.33&-0.18&\onlinecite{Karazhanov}\\ \hline
            c-CdS&  3.44& 1.678 & 0.98&1.97 &0.012&-0.85&-0.83&-0.11&1.34&\onlinecite{Karazhanov} \\ \hline
            c-CdS& 2.2 & 1.058 & 0.36&0.72 &0.02&-0.85&-0.82&-0.5&0.33&\onlinecite{Karazhanov} \\ \hline
&  &  & &  &  &   &  &  &   &   \\ \hline
            w-CdS& 1.71 & 0.62 & -0.203&-0.406 &0.16&-0.98&-1.1&-1.27&-0.61&\onlinecite{Horodyska}\\ \hline
            w-CdS&  1.02& 0.41 & -0.32&-0.64 &0.1&-0.98&-0.98&-1.17&-0.78&\onlinecite{Horodyska} \\ \hline
            w-CdS& 1.09 & 0.34 & -0.43&-0.86 &0.23&-1.05&-1.16&-1.05&-0.96&\onlinecite{Horodyska} \\ \hline
        \end{tabular}
    \end{center}
\end{table}

\end{document}